\documentclass[fleqn,usenatbib]{mnras}
\usepackage{newtxtext,newtxmath}
\usepackage[T1]{fontenc}
\usepackage{graphicx}	
\usepackage{amsmath}	
\usepackage{lineno}
\usepackage{orcidlink}
\graphicspath{{./figures}}

\newcounter{draft}[section]

\title[X-ray Polarization from Hybrid Electrons]{X-ray Polarization of Inverse Compton Scattering by Thermal and Nonthermal Electrons}
\author[B.-T. Wang et al.]{
Bo-Ting Wang$\orcidlink{0000-0001-9342-1485}$,$^{1,2}$
Jirong Mao$\orcidlink{0000-0002-7077-7195}$,$^{1,3,4}$\thanks{E-mail: jirongmao@mail.ynao.ac.cn}
Jie-Ying Liu$\orcidlink{0000-0002-7797-9814}$,$^{1,3,4,5}$
and Fei Xie$\orcidlink{0000-0002-0105-5826}$$^{6,7}$\\
$^{1}$Yunnan Observatories, Chinese Academy of Sciences, Kunming 650216, People’s Republic of China\\
$^{2}$University of Chinese Academy of Sciences, Beijing 100049, People’s Republic of China\\
$^{3}$Center for Astronomical Mega-Science, Chinese Academy of Sciences, 20A Datun Road, Chaoyang District, Beijing 100012, People’s Republic of China\\
$^{4}$Key Laboratory for the Structure and Evolution of Celestial Objects, Chinese Academy of Sciences, Kunming 650216, People’s Republic of China\\
$^{5}$International Centre of Supernovae, Yunnan Key Laboratory, Kunming 650216, People’s Republic of China\\
$^{6}$Guangxi Key Laboratory for Relativistic Astrophysics, School of Physical Science and Technology, Guangxi University, Nanning 530004, People’s Republic of China\\
$^{7}$INAF Istituto di Astrofisica e Planetologia Spaziali, Via del Fosso del Cavaliere 100, 00133 Roma, Italy
}

\date{Accepted 2026 August 6. Received 2026 July 24; in original form 2026 April 9}
\pubyear{\the\year{}}

\begin{document}
\label{firstpage}
\pagerange{\pageref{firstpage}--\pageref{lastpage}}
\maketitle

\begin{abstract}
X-ray emission from accretion-powered astrophysical systems is widely interpreted as inverse Compton (IC) scattering between energetic electrons and soft photons. 
Besides the emitted intensity, the polarization of this radiation provides important information about the physical properties of the electrons involved. We investigate how different electron populations shape both the spectrum and polarization of IC emission in the X-ray band. We consider three electron populations: purely thermal, purely nonthermal power-law, and a hybrid population combining both components. 
We take both numerical simulations and semi-analytic calculations. We first attempt the cases for thermal and nonthermal electrons, respectively. We then focus on the hybrid population, which is expected to be realistic in high-energy object environments.
For the case of hybrid electrons, the scattered emission separates into three energy regimes. At low energies ($\lesssim 0.1$ keV), it is dominated by thermal electrons; at high energies ($\gtrsim 4$ keV), it is governed by the nonthermal component. Between these limits, a transition band ($\sim 0.1\text{–}4$ keV) appears in which both components contribute. 
The degree of polarization varies smoothly across these regimes, and the behavior in the transition band directly traces the relative importance of thermal and nonthermal electrons. 
We further show that for partially polarized seed photons, the scattered polarization scales linearly with the incident polarization while its frequency dependence remains unchanged. These results show that X-ray polarimetry provides a powerful diagnostic of the electron energy distribution in accretion-powered systems.
\end{abstract}

\section{Introduction}
X-ray binaries (XRBs) and active galactic nuclei (AGNs) are among the most powerful X-ray sources in the universe. In these systems, matter accretes onto a compact object, such as a neutron star or a black hole, where gravitational energy is efficiently converted into radiation. As the infalling material descends into the deep gravitational potential well, it is heated to high temperatures and emits radiation over a broad spectral range. In particular, X-ray emission is predominantly produced through inverse Compton (IC) scattering, in which soft seed photons gain energy by interacting with energetic electrons in hot plasma regions located close to the central compact object, such as coronae or inner accretion flows \citep{Eardley1975ApJ, Thorne1975ApJ, Haardt1991ApJ, Haardt1993ApJ}.

Polarization provides a powerful diagnostic for probing the physical processes and geometrical configurations of astrophysical radiation sources. The potential importance of X-ray polarimetry was recognized several decades ago \citep{Rees1975MNRAS}, and early theoretical models explored polarization signatures from accretion disks and scattering media \citep{Lightman1976ApJ}. Measurements of the degree of polarization, and in general of X-ray polarization properties, provide information that is complementary to spectroscopy and timing, enabling discrimination between competing emission mechanisms and source geometries. After a long period of limited observational capability, X-ray polarimetry has recently re-emerged as an 
important research field \citep{Feng2020NatAs}. The launch and successful operation of the Imaging X-ray Polarimetry Explorer (IXPE) has enabled systematic polarization measurements in the 2–8 keV energy band for a wide variety of X-ray sources \citep{2022Weisskopf}. Recent polarization observations of both AGNs and XRBs provide new constraints on the structure and physical conditions of their emitting regions. \citep{Marinucci2022MNRAS, 2023Ingram, Veledina2023ApJ, Ewing2025MNRAS}.

Previous studies have placed important constraints on the geometry of X-ray coronae in accreting systems. Observations such as microlensing and X-ray time-lag measurements have provided estimates of the size, location, and overall configuration of the disk–corona system \citep{Kochanek2004ApJ,Reis2013ApJ}. Beyond these approaches, X-ray polarization offers an independent probe of the coronal environment, as the polarization produced by IC scattering depends not only on the system geometry \citep{Ursini2022MNRAS} but also on the properties of the scattering electrons \citep{Beheshtipour2017ApJ}. Traditional models of IC emission from accreting systems have commonly assumed idealized electron energy distributions, typically either a purely thermal Maxwellian distribution or a nonthermal power-law distribution. Within the framework of the two-phase disk–corona model \citep{Haardt1991ApJ}, polarization properties of Comptonized X-ray emission were investigated by assuming thermal electrons in a hot, optically thin corona above a cold, optically thick disk \citep{Haardt1993MNRAS}. While these prescriptions have provided valuable insights, growing observational and theoretical evidence suggests that electron populations in realistic astrophysical plasmas are unlikely to be described by a single distribution.
Motivated by this, hybrid energy distribution of electrons, consisting of a thermal distribution combined with a nonthermal high-energy tail, has been proposed as a more physically plausible description of hot accretion flows. Early work by \citet{Coppi1999ASPC} introduced hybrid plasmas in which thermal and nonthermal electrons coexist. Subsequent studies developed explicit hybrid distributions featuring a Maxwellian distribution with a power-law tail \citep{2009Giannios_mnras}, and demonstrated their relevance for modeling X-ray spectra from accreting black hole systems. In particular, hybrid electron populations have been invoked to explain the observed hard X-ray spectra of AGN coronae, where nonthermal electrons and pair production can regulate the coronal temperature and produce spectra consistent with observations \citep{Fabian2017MNRAS}.

Hybrid electrons have also been incorporated into more advanced accretion flow models \citep{Davelaar2023MNRAS}, and recent simulations suggest that magnetic reconnection can naturally generate hybrid electron populations in luminous AGN coronae, significantly influencing 
the X-ray emission properties \citep{Liu2024MNRAS}.
We note that the electrons can be relativistic and the thermal temperature defined by $\Theta=kT_e/m_ec^2$ can be larger than 1 \citep{2016ApJ...822...34P}.
In some simulations, the value of $\Theta$ has been usually set to be 10 \citep{Davelaar2023MNRAS}.

It is worth noting that the role of electron energy distributions in shaping X-ray polarization signatures has received comparatively limited attention. Some previous studies have explored the impact of nonthermal electrons on X-ray polarization in accreting systems. \citet{Beheshtipour2017ApJ} included both thermal and nonthermal electron components in their coronal models and investigated the impact of adding a nonthermal energy fraction on the observed X-ray polarization within a general relativistic ray-tracing framework. More recently, \citet{Moscibrodzka2022ApJS} examined IC scattering from electrons described by different assumed energy distribution functions—including thermal, power-law, hybrid and $\kappa$ distributions—within a numerical scattering framework. In that work, hybrid electron distributions were treated as representative inputs for testing the consistency and accuracy of the scattering kernel and its polarization properties, with an emphasis on validating the numerical scheme against analytic expectations, rather than on a systematic exploration of how hybrid electron parameters shape polarization signatures. 
In this paper, we focus on the intrinsic polarization produced by IC scattering from hybrid electron energy distributions. We systematically examine how the transition between thermal and nonthermal components, regulated by the parameters of the hybrid distribution, influences the degree of polarization across different energy bands.

The intrinsic polarization calculations presented in this work are carried out using a Monte Carlo (MC) treatment of IC scattering, following the approach described and formalized by \citet{Krawczynski2012ApJ}. 
In addition, semi-analytic (SA) expressions for the frequency dependence of IC polarization, derived by \citet{Krawczynski2012ApJ} based on the formalism of \citet{Bonometto1970A&A}, are employed as a reference for comparison and validation of the numerical results. For completeness, the numerical scheme and the SA formulation are summarized in Appendix~\ref{appendixa} and Appendix~\ref{appendixb}, respectively. The formulation adopted in this work, based on the SA approach of \citet{Bonometto1970A&A}, provides a valid description of the polarization properties of the scattered radiation. More generally, a covariant treatment of polarized Compton scattering has been developed by \citet{nagirner1993} and \citet{poutanen1993A&A}. Their general scattering matrix offers an equivalent description of the same physical process and, in appropriate limits, reduces to the expressions derived by \citet{Bonometto1970A&A} for specific electron distributions, such as the power-law case. We therefore regard the present approach and the covariant formalism as physically equivalent. 
The formulation developed by \citet{nagirner1993} has also been renewed for some applications in recent studies, for example, in the polarized radiative transfer calculations of \citet{Krawczynski2022Sci}.

This paper is organized as follows. In Section~\ref{Electron energy distribution}, we describe the electron energy distributions considered in this work and the sampling procedures adopted in the numerical calculations. The polarization characteristics of IC scattering by hybrid electron populations are presented in Section~\ref{Results}. Finally, we discuss the implications of our findings and outline prospects for future work in Section~\ref{Discussion and Conclusions}.

\section{Scattering Geometry and Polarization Convention}
\begin{figure}
    \centering
    \includegraphics[width=\columnwidth]{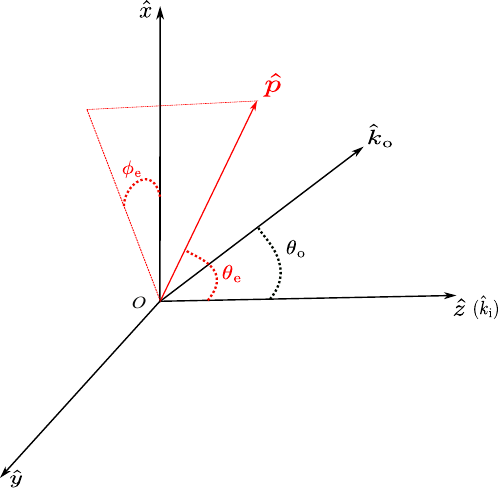}
    \hfill
    \includegraphics[width=\columnwidth]{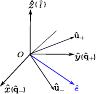}
    \caption{
    \textit{Top panel}: Spherical polar coordinate system used to describe Compton scattering in the plasma frame. The incident photon initially travels along the $\hat{z}$-axis. Directions are defined as follows: $\hat{p}$ is the direction of the electron, $\hat{k}_{\rm i}$ is the direction of the incident photon and $\hat{k}_{\rm o}$ is the direction of the scattered photon. The angle shown in the figure is defined as follows: $\theta_{\rm o}\equiv \sphericalangle(\hat{k}_{\rm i},\hat{k}_{\rm o})$. 
    \textit{Bottom panel}: Schematic illustration of the Stokes-vector geometry. The photon propagation direction is defined by $\hat{l}$, which is aligned with the $\hat{z}$-axis. The polarization basis vectors $\hat{u}_{-}$ and $\hat{u}_{+}$ lie in the plane perpendicular to the photon propagation direction (the $\hat{x}$–$\hat{y}$ plane). $\hat{e}$ denotes the electric-field vector.
    }
    \label{fig:fig1}
\end{figure}

The calculations are defined in the plasma frame (PF). The coordinate system in the top panel of Figure~\ref{fig:fig1} has $\hat{\mathbf{k}}_{\rm i}=\hat{\mathbf{z}}$, where $\hat{\mathbf{k}}_{\rm i}$ is the incident-photon direction. The electron direction $\hat{\mathbf{p}}$ and scattered-photon direction $\hat{\mathbf{k}}_{\rm o}$ are specified by $(\theta_{\rm e},\phi_{\rm e})$ and $(\theta_{\rm o},\phi_{\rm o})$, respectively. The polar angles $\theta_{\rm e}$ and $\theta_{\rm o}$ are measured from $\hat{\mathbf{z}}$. The azimuthal angles $\phi_{\rm e}$ and $\phi_{\rm o}$ are measured about $\hat{\mathbf{z}}$ from $\hat{\mathbf{x}}$. In this coordinate system, $\theta_{\rm o}=\sphericalangle(\hat{\mathbf{k}}_{\rm i},\hat{\mathbf{k}}_{\rm o})$. The configuration shown in Figure~\ref{fig:fig1} has $\phi_{\rm o}=0$.

The light carries the dimensionless Stokes vector $\mathbf{s}=(i,q,u)$. The component $i$ gives its statistical weight, and $q$ and $u$ specify its linear polarization state. For a photon propagating along the local $\hat{\mathbf{z}}$-axis, $q/i=+1$ and $q/i=-1$ correspond to electric fields parallel to $\hat{\mathbf{y}}$ and $\hat{\mathbf{x}}$, respectively. The states $u/i=+1$ and $u/i=-1$ correspond to electric fields parallel to $\hat{\mathbf{u}}_{+}=(\hat{\mathbf{y}}-\hat{\mathbf{x}})/\sqrt{2}$ and $\hat{\mathbf{u}}_{-}=(\hat{\mathbf{y}}+\hat{\mathbf{x}})/\sqrt{2}$, respectively. The degree of linear polarization is $\Pi=\sqrt{q^2+u^2}/i$. The detailed MC simulation of the polarization in IC scattering is given in Appendix~\ref{appendixa}.

\section{Electron energy distribution}\label{Electron energy distribution}

\begin{figure}
    \centering
    \includegraphics[width=\columnwidth]{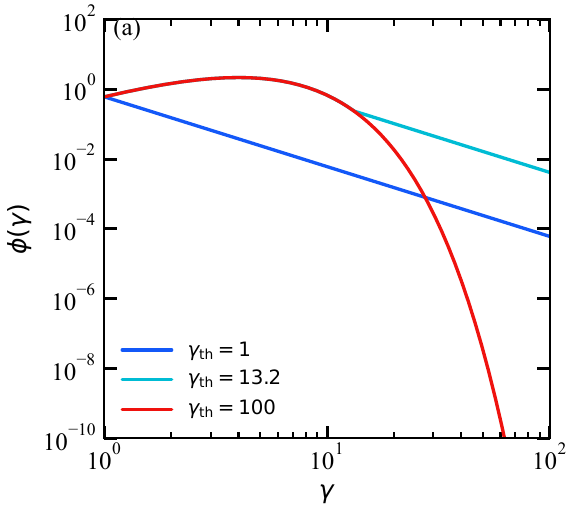}
    \hfill
    \includegraphics[width=\columnwidth]{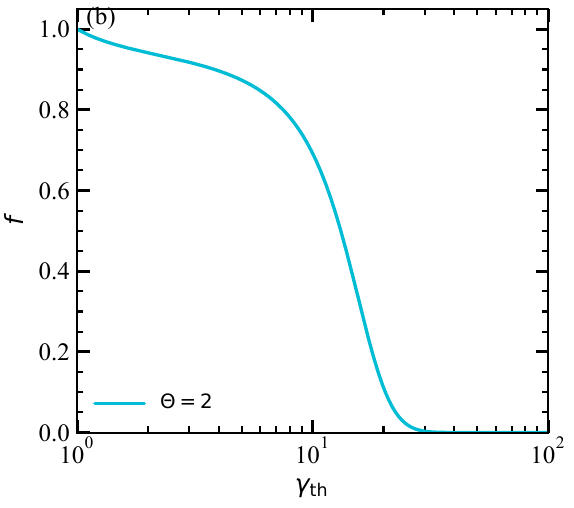}
    \caption{
    (a) The unnormalized hybrid electron energy distributions $\phi(\gamma)$ for $\Theta=2$. The curves represent different transition Lorentz factors, $\gamma_{\rm th}$, representing pure power-law (blue), hybrid (cyan), and pure thermal (red) distributions. 
    (b) The nonthermal electron energy fraction as the function of $\gamma_\text{th}$ for $\Theta=2$. The power-law index for the nonthermal tail is fixed at $p_{\text{hybrid}}=2$ for all cases.
    }
    \label{fig:fig2}
\end{figure}

We adopt the hybrid energy distribution function (DF) of electrons consisting of a thermal Maxwell–Jüttner distribution smoothly joined to a non-thermal power-law tail. The unnormalized shape function is defined as \citep{2009Giannios_mnras}
\begin{equation}\label{hybrid dist}
\phi(\gamma)=
\begin{cases}
\gamma^{2}\exp(-\gamma/\Theta), & \gamma\le\gamma_{\rm th},\\
\gamma_{\rm th}^{2}\exp(-\gamma_{\rm th}/\Theta)\left(\dfrac{\gamma}{\gamma_{\rm th}}\right)^{-p_{\mathrm{hybrid}}}, & \gamma>\gamma_{\rm th},
\end{cases}    
\end{equation}
where $\Theta=kT_{\rm e}/m_{\rm e}c^{2}$ is the dimensionless temperature, $p_{\text{hybrid}}$ is the power-law index, and $\gamma_{\text{th}}$ marks the transition between the thermal and non-thermal components. We focus on the relativistic regime, where the condition $\Theta > 1$ should be satisfied. The unnormalized hybrid energy distribution of electrons $\phi(\gamma)$ for $\Theta = 2$ as an example is shown in Figure \ref{fig:fig2}(a). The normalized electron distribution is written as
\begin{equation}
    f(\gamma)= C_{\rm e}\,\frac{\phi(\gamma)}{2\Theta^{3}}, \
\int_{\gamma_{1}}^{\gamma_{th}} f(\gamma)_\text{thermal}\,{\rm d}\gamma+\int_{\gamma_{th}}^{\gamma_{2}} f(\gamma)_\text{nonthermal}\,{\rm d}\gamma = 1 ,
\end{equation}
with the normalization constant obtained from $C_{\rm e}
=
\left[
\int_{\gamma_{1}}^{\gamma_{2}}
\frac{\phi(\gamma)}{2\Theta^{3}}\,{\rm d}\gamma
\right]^{-1}.$ The parameters are the dimensionless thermal temperature $\Theta$
, the tail power-law index $p_{\text{hybrid}}$, and the transition Lorentz factor $\gamma_{\rm th}$.

To generate MC samples from the hybrid distribution, we treat the thermal and non-thermal components separately. Thermal electrons ($ \gamma_{1}\le\gamma \le \gamma_{\rm th}$) are sampled using the Maxwell–Jüttner procedure introduced by \citet{Canfield1987ApJ},  a method also utilized in the {\sc grmonty}\footnote{\url{https://github.com/black-hole-group/grmonty}} and {\sc $\kappa$monty}\footnote{\url{https://github.com/jordydavelaar/kmonty}} codes \citep{Dolence_2009,Davelaar2023MNRAS}. For electrons in the non-thermal regime ($\gamma_{\rm th} \le\gamma \le \gamma_{2}$), the cumulative distribution function (CDF) is given by
\begin{equation}\label{pw_cdf}
F(\gamma)
= \int_{\gamma_{th}}^{\gamma}
\frac{(p_{\text{hybrid}}-1)\,\gamma^{-p_{\text{hybrid}}}}{\gamma_{th}^{\,1-p_{\text{hybrid}}}-\gamma_{2}^{\,1-p_{\text{hybrid}}}}
\, \mathrm{d}\gamma
= \frac{\gamma_{th}^{\,1-p_{\text{hybrid}}}-\gamma^{\,1-p_{\text{hybrid}}}}
{\gamma_{th}^{\,1-p_{\text{hybrid}}}-\gamma_{2}^{\,1-p_{\text{hybrid}}}} .
\end{equation}
Since this CDF is analytically invertible, we employ the inverse transform sampling method. We first draw a random variate $u$ from a uniform distribution over the interval $[0, 1]$. By setting $F(\gamma) = u$ and inverting Equation (\ref{pw_cdf}), we determine the Lorentz factor $\gamma$ as follows:
\begin{equation}
\gamma=
\left[
\gamma_{th}^{\,1-p_{\text{hybrid}}}(1-u)
+
\gamma_{2}^{\,1-p_{\text{hybrid}}} u
\right]^{1/(1-p_{\text{hybrid}})}.
\end{equation}
Through the sampling procedure described above, the hybrid energy distribution of electrons is effectively sampled.

To investigate the fraction of electron energy associated with the thermal and non-thermal components, we follow the prescription of \citet{2009Giannios_mnras} and introduce a parameter $f$, defined as
\begin{equation}\label{energy fraction}
    f =\frac{\displaystyle\int_{\gamma_{\rm th}}^{\infty} \gamma\, \phi(\gamma)\, d\gamma}
     {\displaystyle\int_{1}^{\infty} \gamma\, \phi(\gamma)\,d\gamma },
\end{equation}
$\phi(\gamma)$ is determined by Equation (\ref{hybrid dist}). In Figure \ref{fig:fig2}(b), we show the parameter $f$ as a function of the conjunctive Lorentz factor $\gamma_{\rm th}$. The parameter $f$ represents the fraction of energy carried by non-thermal electrons. When $\gamma_{\rm th} = 1$, we obtain $f = 1$, indicating that the total electron energy is entirely contributed by the non-thermal component. As $\gamma_{\rm th}$ increases, the thermal component progressively dominates the total energy.

\section{Results}\label{Results}
\subsection{Validation of the Sampling Procedure}
Before presenting the results, it is worth noting that the MC calculations of IC polarization in this work follow the framework of \citet{Krawczynski2012ApJ}. As an analytic reference, the SA expressions developed by \citet{Bonometto1970A&A} are employed. The results below are structured to validate the electron distribution sampling.

\begin{figure}
    \centering
    \includegraphics[width=\columnwidth]{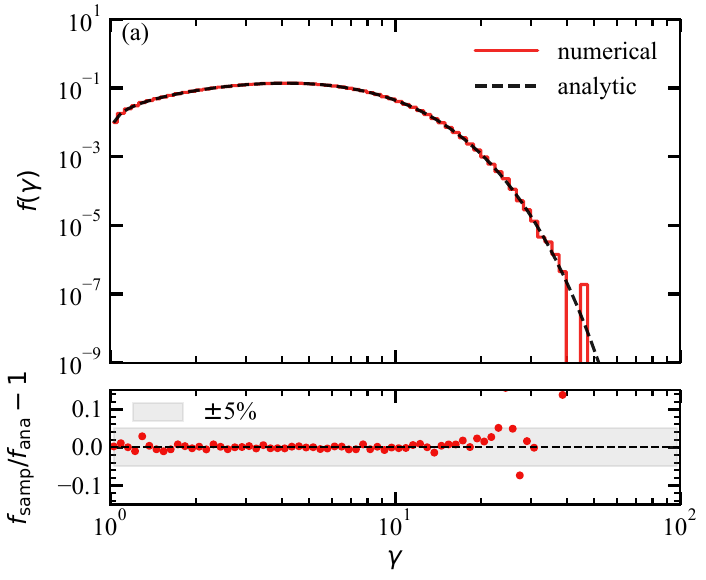}
    \hfill
    \includegraphics[width=\columnwidth]{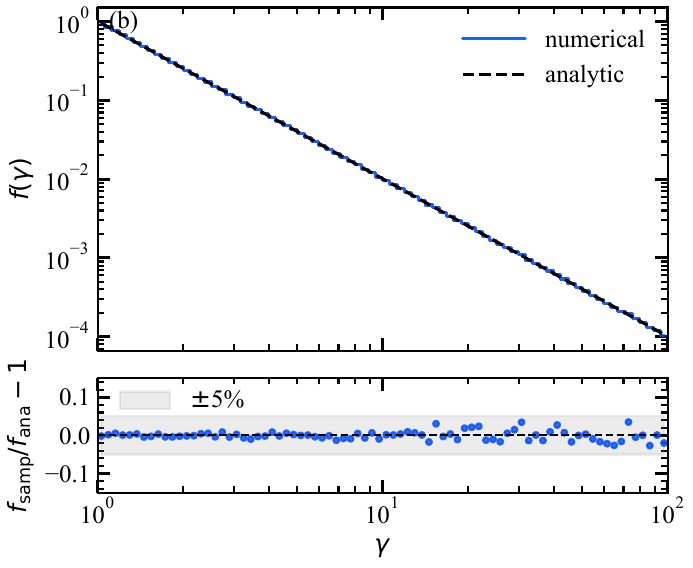}
    \vspace{2mm}
    \includegraphics[width=\columnwidth]{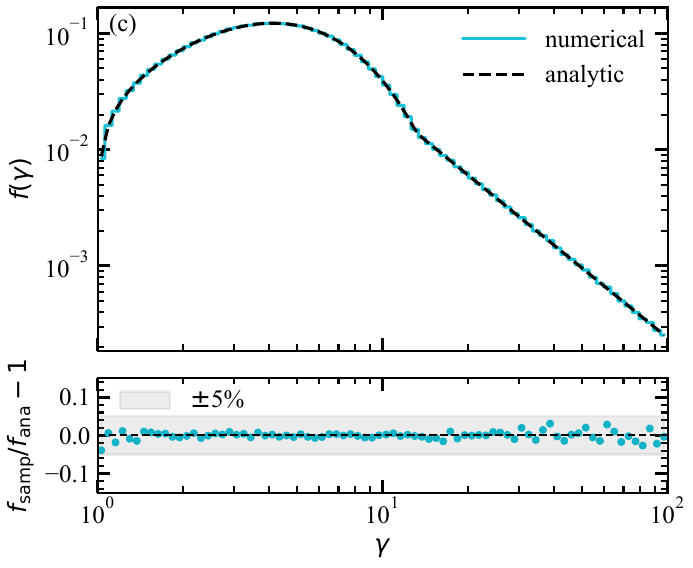}
    \caption{
    Comparison between the analytic electron energy distribution $f(\gamma)$ and MC sampling, with all distributions computed over the Lorentz factor interval $\gamma \in [1,100]$. 
    (a) Thermal Maxwell-Jüttner distribution with $\Theta = 2$. 
    (b) Power-law distribution with index $p_{\text{hybrid}} = 2$. 
    (c) Hybrid distribution combining a thermal distribution with $\Theta = 2$ and a nonthermal tail with $p_{\text{hybrid}}=2$, with a transition at $\gamma_\text{th}=13.2$.
    }
    \label{fig:fig3}
\end{figure}

We validate the sampling method by comparing the MC results with analytic electron energy distributions for three representative cases: a thermal Maxwell--Jüttner distribution ($\Theta = 2$), a nonthermal power-law distribution with index $p_\text{hybrid} = 2$\footnote{Here we adopt the value used in \citet{Dreyer2021ApJ}.}, and a hybrid distribution consisting of a thermal distribution ($\Theta = 2$) and a nonthermal tail ($p_\text{hybrid} = 2$), with a transition at $\gamma_\text{th}=13.2$\footnote{Here, we choose $\gamma_{\rm th} = 13.2$, according to Equation (\ref{energy fraction}) indicates that this value produces a hybrid electron energy distribution with equal contributions from the thermal and nonthermal components.}. We generate $2\times10^{6}$ electrons and compare the sampled distribution with its analytical form. The results are shown in Figure \ref{fig:fig3}. For all three cases considered above, the relative difference between the analytical and sampled distributions remains nearly zero. The results indicate that for a relatively large value of $\gamma$, the DF contains a small number of electrons and is therefore dominated by MC noise.

\subsection{Polarization Signatures of IC Scattering for Different Electron Distributions}
Before investigating IC scattering by hybrid electron energy distributions, we first consider the scattering of monoenergetic photons by isotropic, monoenergetic electrons.
Although the electrons are usually considered to be isotropic, the cross-section of the scattering is anisotropic (See Appendix \ref{appendixa} in detail). 
Figure \ref{fig:fig4} shows the frequency dependence of the scattered intensity and the degree of polarization for several representative electron Lorentz factors, $\gamma_0 = 2,5,10,20$ and $100$.

\begin{figure}
\includegraphics[width=\columnwidth]{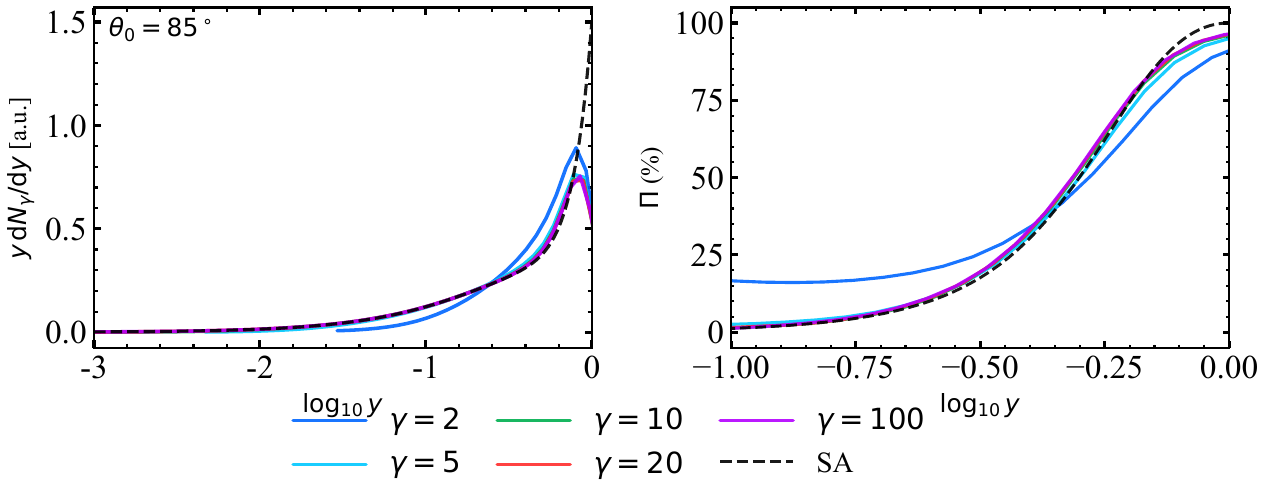}
\caption{IC intensity (left) and the degree of polarization (right) for an isotropic, monoenergetic electron population with Lorentz factors $\gamma = 2, 5, 10, 20$, and $100$, scattering a monoenergetic, unidirectional photon beam with frequency $\omega=5\times10^{14}$ Hz and initial Stokes parameters ${\bf s}=(1,1,0)$. The results are shown for photons emitted into the direction $(\theta_{\rm o},\phi_{\rm o})=(85^\circ,0^\circ)$. Solid curves denote numerical results, while dashed curves show the corresponding SA predictions.}
\label{fig:fig4}
\end{figure}

We introduce the dimensionless scattered frequency $y \equiv \omega_{\rm o}/\omega_{\rm max}$, where $\omega_{\rm o}$ is the frequency of the scattered photon and $\omega_{\rm max}$ denotes the maximum scattered photon frequency. We show the differential photon number ${\rm d}N_\gamma/{\rm d}y$  (see Appendix \ref{appendixb} for a detailed definition), which represents the intensity kernel for monoenergetic electrons with Lorentz factor $\gamma_0$.
For all the cases, the scattered intensity increases toward the maximum scattered photon frequencies, indicating the increasing contribution from photons scattered to higher energies. 
The degree of polarization increases monotonically with the scattered photon frequency and approaches high values ($\Pi \gtrsim 80\%$) as the photons near the maximum scattered frequency.
Although the overall trends are similar for different electron energies, systematic differences emerge at low frequencies of scattered photons. 
Specifically, in this low-frequency regime, lower-energy electrons tend to produce a higher degree of polarization than higher-energy electrons, indicating differences in the scattering characteristics associated with different electron energies. 
The results for different viewing angles are presented in Figure~\ref{fig:fig4_app_1}.

\begin{figure}
    \centering
    \includegraphics[width=\columnwidth]{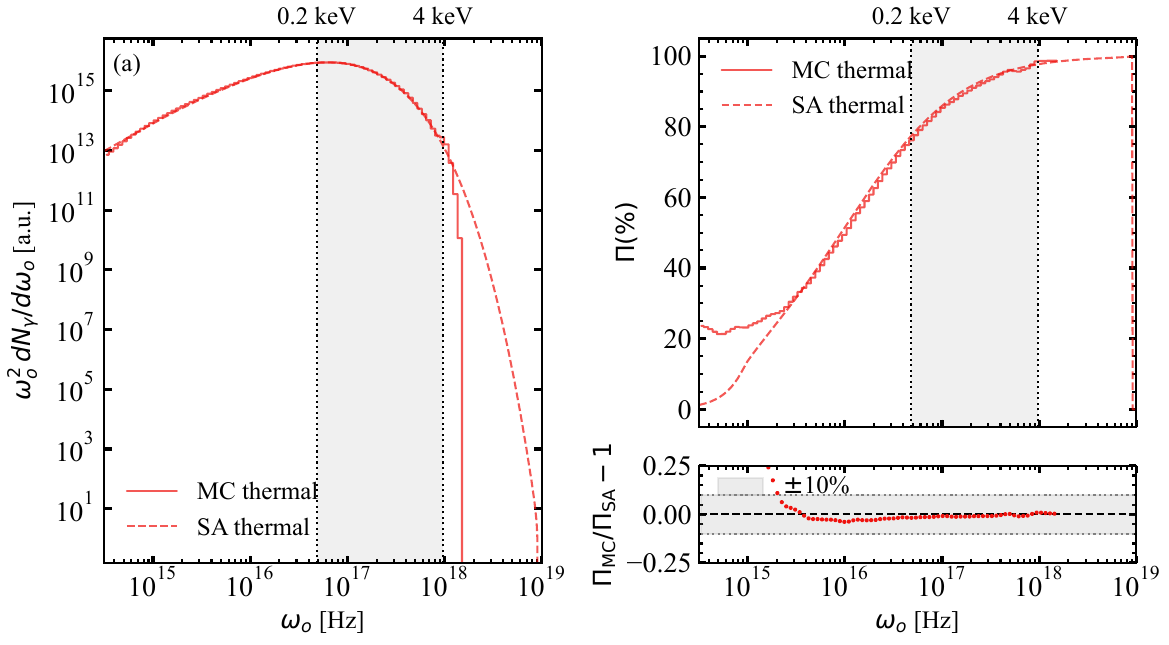}
    \hfill
    \includegraphics[width=\columnwidth]{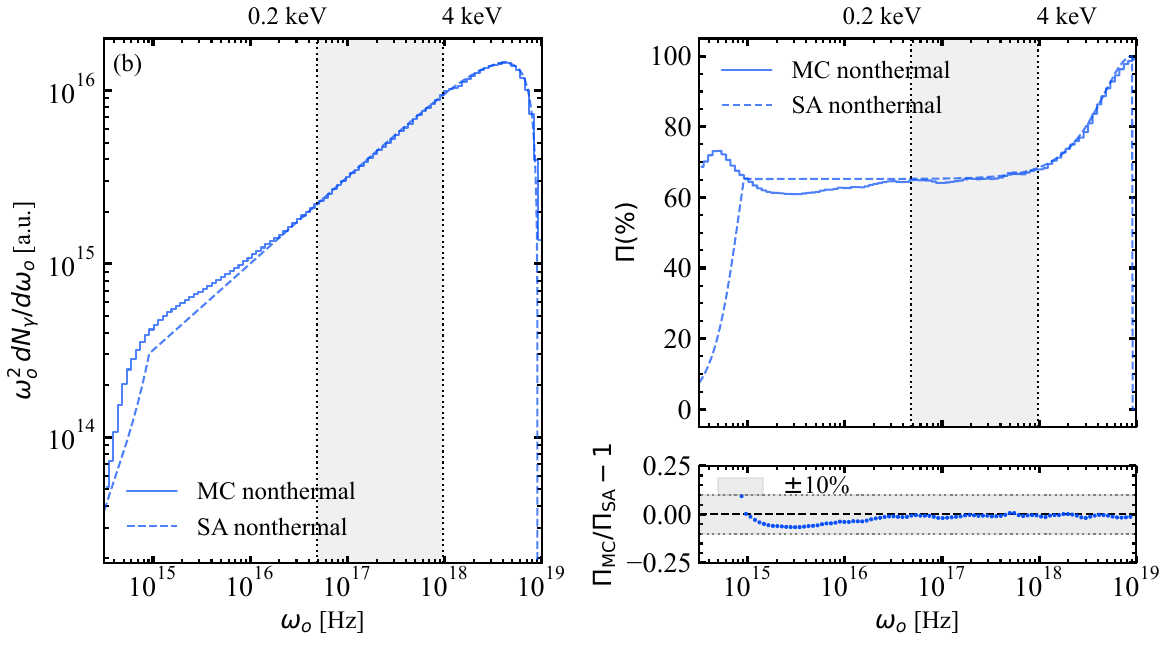}
    \caption{
    IC intensity (left) and the degree of polarization (right) for the isotropic electron energy distributions, calculated over the Lorentz factor range $\gamma \in [1,100]$, representing the scattering of the monoenergetic beam ($\omega=5\times10^{14}$ Hz, ${\bf s}=(1,1,0)$) into $(\theta_{\rm o},\phi_{\rm o})=(85^\circ,0^\circ)$. (a) Thermal Maxwell-Jüttner distribution with $\Theta = 2$. The dashed and solid lines represent the SA and numerical results, respectively. (b) Power-law distribution with index $p_{\text{hybrid}} = 2$. The dashed and solid lines denote the SA and numerical results, respectively. The vertically shaded region marks the 0.2–4 keV energy range.}
    \label{fig:fig5}
\end{figure}

We next examine two single-component scenarios, i.e., a pure relativistic thermal (Maxwell–Jüttner) distribution and a pure power-law distribution. Figure~\ref{fig:fig5} shows the frequency dependence of the IC intensity and the degree of polarization for the two single-component electron distributions. The calculations are performed for isotropic electrons in the Lorentz factor range $\gamma \in [\gamma_1,\gamma_2]$, where $\gamma_1 = 1$ and $\gamma_2 = 100$. We consider the scattering of a monoenergetic incident photon beam with frequency $\omega = 5 \times 10^{14} {\rm Hz}$ into the direction $(\theta_{\rm o},\phi_{\rm o}) = (85^\circ,0^\circ)$. Panel (a) of Figure~\ref{fig:fig5} shows the results for a purely thermal electron distribution with temperature $\Theta = 2$, while panel (b) shows the results for a pure power-law distribution with index $p_{\rm hybrid} = 2.0$.

For a purely thermal electron population, the IC spectrum exhibits a characteristic broad peak followed by a sharp high-frequency cutoff, indicating the rapid decline of the thermal electron distribution at large Lorentz factors. At low scattered frequencies, emission is produced at small values of $y$, and the degree of polarization remains modest. As the scattered frequency increases toward the spectral turnover, the degree of polarization increases monotonically and reaches a high polarization level as the scattered photon frequency approaches its maximum value for the thermal electron population. Near the high-frequency cutoff, as $\omega_{\rm o}$ approaches its maximum value for the thermal population, the degree of polarization $\Pi$ tends toward $\sim100\%$, consistent with emission produced at scattered photon frequencies close to the maximum ($y\rightarrow1$). The numerical results are consistent with SA predictions over the majority of the frequency range. The lower panel presents the relative deviation between the numerical and SA results.
Deviations are observed at the low frequencies, where scattering is dominated by mildly relativistic electrons and the SA approximations become less accurate. This discrepancy arises because the SA formalism derived by \citet{Bonometto1970A&A} is valid for electron Lorentz factors $\gamma \gtrsim 10$ \citep{Krawczynski2012ApJ}. Therefore, when the low-frequency emission is produced mainly by scattering off mildly relativistic electrons ($\gamma < 10$), the SA and numerical results exhibit 
pronounced deviations. 
At the high-frequency end of the spectrum, noticeable deviations between the numerical and SA results are also observed.
These deviations are mainly attributed to the statistical uncertainty of the MC calculation rather than the accuracy of the SA approximation.
In this regime, the scattered photons are produced primarily by electrons with large Lorentz factors ($\gamma \gtrsim 40$), where the Maxwell-Jüttner distribution has a low particle number density.
Therefore, the effective number of sampled electrons contributing to this part of the spectrum is limited, leading to increased statistical fluctuations in the MC estimate of the intensity.


For a pure power-law electron distribution, the IC emission spectrum exhibits three distinct regimes: (i) a low-frequency region corresponding to $\gamma_1 < \gamma_{\min}$; (ii) a power-law region beginning at $\gamma_{\min} \gtrsim \gamma_1$; and (iii) a high-frequency region dominated by electrons with Lorentz factors approaching $\gamma_2$. Here $\gamma_{\min}$ is defined as the minimum Lorentz factor required for a seed photon of frequency $\omega$ to be scattered into a photon of observed frequency $\omega_{\rm o}$ (see Appendix~\ref{appendixb} for the explicit expression). In the low-frequency regime, the degree of polarization is relatively small because the emission is produced at low values of $y$. Notably, in this regime, the numerical results show a comparatively large deviation from the SA predictions, indicating the same limitations of the SA formalism discussed above. Within the intermediate power-law regime, the degree of polarization is essentially independent of the frequency of scattered photon, $\omega_{\rm o}$. Conversely, in the high-frequency limit where $\omega_{\rm o} \rightarrow\omega_{\max}(\gamma_2)$, corresponding to $y\rightarrow 1$, the degree of polarization increases toward $\Pi = 100\%$.

\begin{figure}
    \centering
    \includegraphics[width=\columnwidth]{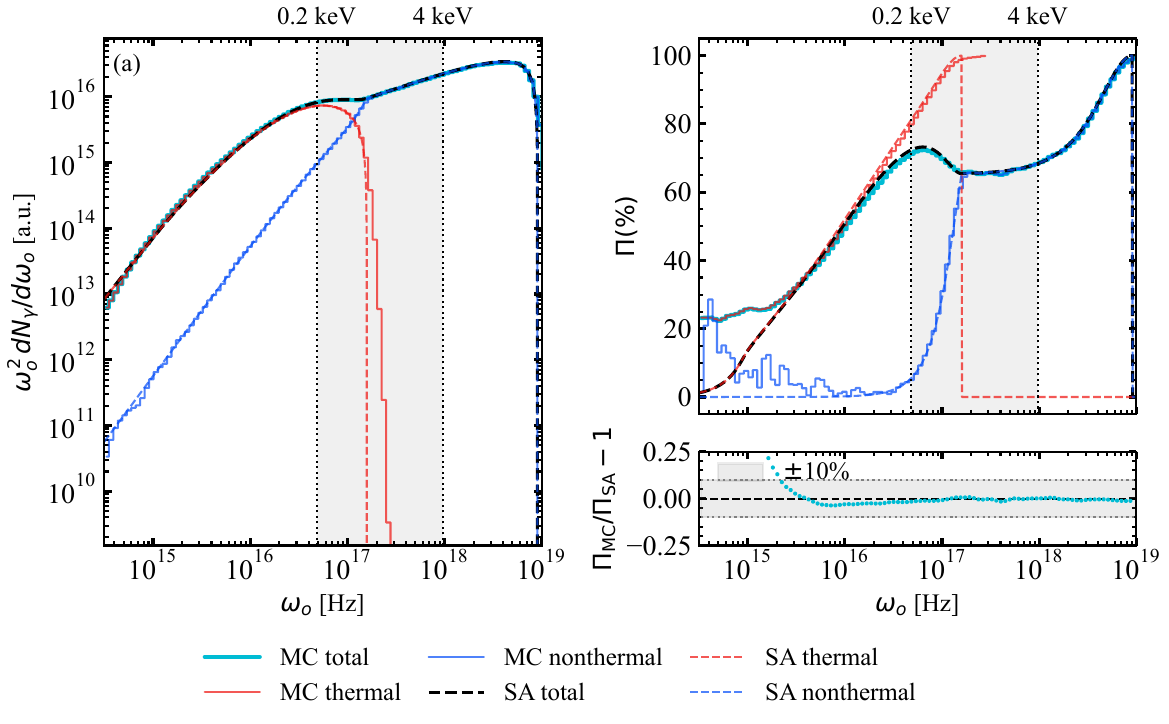}
    \hfill
    \includegraphics[width=\columnwidth]{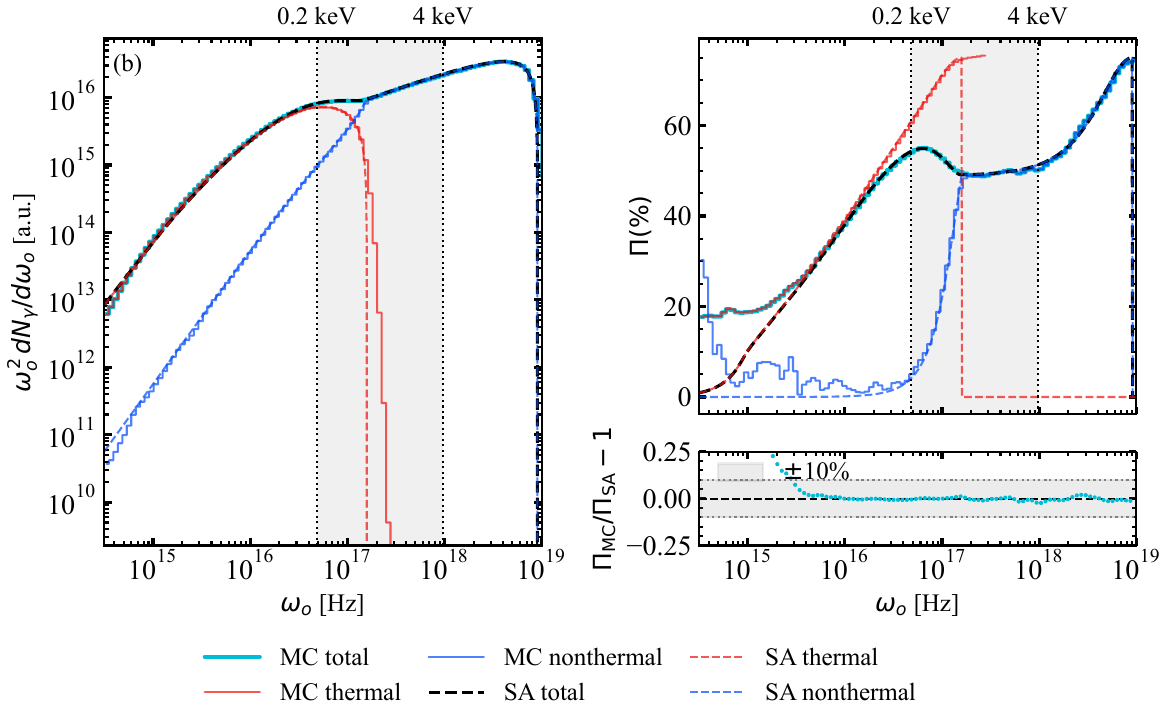}
    \vspace{2mm}
    \includegraphics[width=\columnwidth]{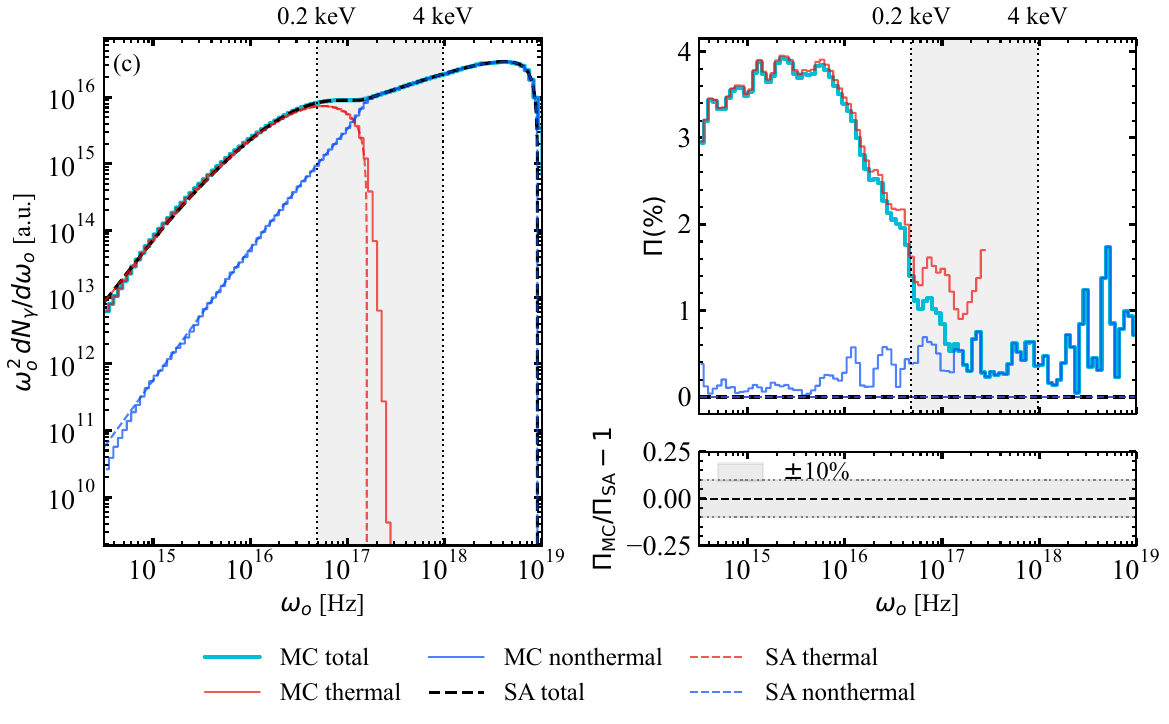}
    \caption{
    IC intensity (left) and the degree of polarization (right) for the isotropic hybrid electron energy distributions with $p_{\text{hybrid}}=2$, $\gamma_{\rm th}=13.2$, $\Theta = 2$ and $\gamma\in[1,100]$, representing the scattering of the monoenergetic beam ($\omega=5\times10^{14}$ Hz) into $(\theta_{\rm o},\phi_{\rm o})=(85^\circ,0^\circ)$. Numerical and SA results are shown for the total (cyan), thermal (red), and non-thermal (blue) components. 
    (a) ${\bf s}=(1,1,0)$. 
    (b) ${\bf s}=(1,0.75,0)$. 
    (c) ${\bf s}=(1,0,0)$. The vertically shaded region marks the 0.2–4 keV energy range.
    }
    \label{fig:fig6}
\end{figure}

We now turn to the case of the hybrid energy distribution of electrons, in which the thermal distribution and nonthermal tail conjunctively determine the spectral and polarization properties of the scattered emission. The calculations are performed for isotropic electrons in the range $\gamma \in [\gamma_1,\gamma_2]$, adopting $\gamma_1 = 1$ and $\gamma_2 = 100$. In panel (a) of Figure~\ref{fig:fig6}, we show the frequency dependence of the scattered intensity and the degree of polarization for a hybrid electron distribution with $\Theta=2$, $p_{\text{hybrid}}=2$, and $\gamma_{\rm th}=13.2$, assuming completely polarized seed photons and a viewing angle of $\theta_{\rm o}=85^\circ$. For the hybrid energy distribution of electrons, the IC emission exhibits distinct frequency-dependent behaviors indicating the combined contributions of the thermal distribution and the nonthermal tail. We have three results. (1) At low scattered frequencies, $\omega_{\rm o} \lesssim 10^{16.5}$ Hz ($\lesssim 0.1$ keV), the emission is dominated by thermal electrons. In this regime, the total intensity closely follows the thermal component, and the degree of polarization remains moderate, similar to the purely thermal case. The contribution from the nonthermal tail is negligible at these frequencies. (2) At intermediate frequencies, $10^{16.5} \lesssim \omega_{\rm o} \lesssim 10^{18}$ Hz ($\sim 0.1\text{–}4$ keV), the contributions from thermal and nonthermal electrons become comparable. The spectrum exhibits a smooth transition between the two components, while the degree of polarization rises gradually from the value characteristic of the thermal component toward the higher value associated with nonthermal scattering. The transition frequency band serves as an effective diagnostic of the hybrid electron population because the total polarization is highly sensitive to the relative contributions and polarization properties of the thermal and nonthermal components.
The behavior observed in the transition regime can be understood by noting that the total polarization is effectively a weighted combination of the thermal and nonthermal contributions, with the weights given by their respective intensities. In the transition regime, although nonthermal electrons start to contribute to the scattered intensity (\(I_{\rm nth} > 0\)), their associated degree of polarization at a given scattered frequency can be lower than that of thermal electrons (\(\Pi_{\rm nth} < \Pi_{\rm th}\)). 
As a result, the addition of the nonthermal component causes the total degree of polarization to drop below the value corresponding to a purely thermal electron population.
(3) At high frequency regime, $\omega_{\rm o} \gtrsim 10^{18}$ Hz ($\gtrsim 4$ keV), the emission is dominated by the nonthermal power-law electrons. In this regime, the total intensity becomes dominated by the nonthermal component, and the degree of polarization approaches the high-frequency plateau observed in the pure power-law case. As the scattered photon frequency approaches its maximum value, the polarization increases toward $\Pi \sim 100\%$. For unpolarized seed photons scattering off relativistic electrons, theory predicts that no net polarization should be produced \citep{Bonometto1970A&A}. Our numerical results are consistent with this expectation. As shown in panel (c) of Figure~\ref{fig:fig6}, the small residual polarization apparent is attributable to MC statistical fluctuations rather than physical effects.

We assumed the seed photons to completely polarized and obtained the results shown in panel (a) of Figure~\ref{fig:fig6}. When partially polarized seed photons are considered (see panel (b) of Figure~\ref{fig:fig6}), the scattered polarization follows a simple linear scaling. Specifically, if scattering reduces the degree of polarization of a completely polarized beam to $\Pi_1$, an incident beam with an initial polarization $\Pi_2$ emerges with a degree of polarization $\Pi = \Pi_1 \Pi_2$. Consequently, the frequency dependence is preserved and only the overall amplitude is rescaled. This behavior is consistent with the well-established result for IC scattering by purely nonthermal electrons: the process generates no polarization from initially unpolarized photons but reduces the degree of polarization of the polarized incident photons \citep{Bonometto1970A&A}. Our results show that the same scaling applies to hybrid electron populations. We verify this scaling in the 0.1–4~keV band as an example. For completely polarized seed photons, we obtain $\Pi_1=67.9\%$, consistent with the SA prediction of $67.1\%$. For partially polarized seed photons with $\Pi_2=75\%$, the scaling predicts $\Pi \simeq 0.679\times 0.75=50.9\%$. The numerical result, $\Pi=51.4\%$, agrees with this expectation and is consistent with the SA prediction of $50.3\%$. Importantly, this scaling preserves the frequency-dependent polarization signatures associated with the hybrid electron distribution, affecting only the overall polarization amplitude. As shown in Figure \ref{fig:fig6}, the intensity of IC radiation remains independent of the degree of polarization of the incident photon \citep{Bonometto1970A&A, Moscibrodzka2022ApJS}. 

For the X-ray band, for example, in the energy band of 2–8 keV, for the completely polarized seed photons, our model predicts a band-averaged polarization of $\Pi_{\rm 2-8 keV}\approx67\%$. 
We consider the seed photons from an accretion disk with an intrinsic polarization of $\Pi_{\rm disk}\approx8\%$ \citep{Kushwaha2023MNRAS,rawat2023detection,Ratheesh2024ApJ} will retain about $\Pi_{\rm IC}\approx0.67\times 8\%\approx5\%$ polarization after undergoing inverse-Compton scattering by the hybrid energy distribution of electrons.

\subsection{Effects of $\gamma_{\rm th}$ on the IC Polarization}
\begin{figure}
\includegraphics[width=\columnwidth]{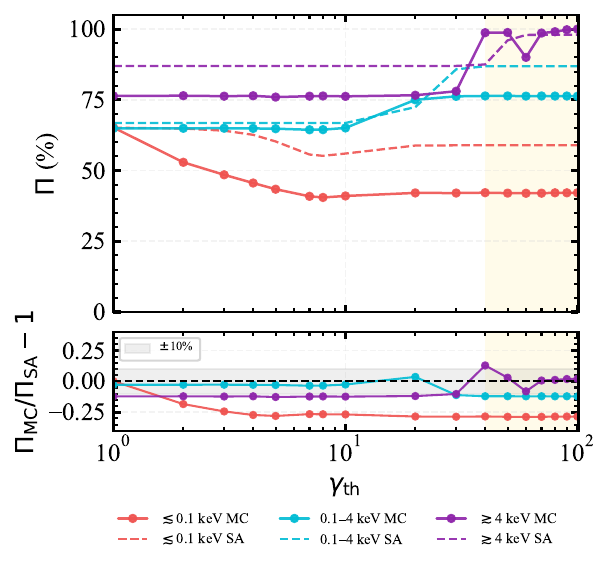}
\caption{IC polarization as a function of the hybrid-electron transition Lorentz factor $\gamma_{\rm th}$ for a hybrid distribution with $\gamma\in[1,100]$, $p_{\text{hybrid}}=2$, and $\Theta=2$, scattering a monoenergetic beam ($\omega=5\times10^{14}$ Hz, ${\bf s}=(1,1,0)$) into $(\theta_{\rm o},\phi_{\rm o})=(85^\circ,0^\circ)$. The degree of polarization $\Pi$ is shown for three energy bands: $\lesssim 0.1$ keV (red), $0.1–4$ keV (cyan), and $\gtrsim 4$ keV (purple). Solid symbols denote numerical results, while dashed curves show the corresponding SA predictions.}
\label{fig:fig7}
\end{figure}

The transition Lorentz factor $\gamma_{\rm th}$ regulates the relative contributions of the thermal and nonthermal components in the hybrid electron energy distribution. 
Figure~\ref{fig:fig7} shows the dependence of the band-averaged degree of polarization on $\gamma_{\rm th}$ across different energy bands, illustrating how the polarization varies with this control parameter. 
The low-energy band exhibits a relatively low degree of polarization with only a weak dependence on $\gamma_{\rm th}$, indicating that the scattered emission in this band is largely insensitive to variations in the nonthermal contribution. 
In contrast, the high-energy band maintains a consistently high degree of polarization and gradually approaches a stable value as $\gamma_{\rm th}$ increases, indicating the dominant role of high-energy electrons in this regime. 
The significant dependence on $\gamma_{\rm th}$ appears in the intermediate energy band, where the degree of polarization increases with increasing $\gamma_{\rm th}$. 
This behavior directly traces the evolving relative contributions between the thermal and nonthermal components in the hybrid distribution, indicating that this band provides an effective diagnostic.
The corresponding frequency-dependent intensity and polarization spectra for different values of $\gamma_{\rm th}$ are presented in Figure~\ref{fig:fig7_app_1},
which illustrate how the spectral shape and polarization evolution change as the transition Lorentz factor varies.
The lower panel presents the relative differences between the numerical and SA results. 
For the intermediate energy band (0.1–4 keV), the two approaches show good agreement across most of the parameter space, with deviations typically confined within $\pm10\%$. 
At lower energies ($\lesssim 0.1$ keV), larger deviations are observed. In this regime, scattering is dominated by mildly relativistic electrons, for which the SA approximations become less accurate, leading to systematic differences between the numerical and SA results \citep{Krawczynski2012ApJ}. 
At higher energies ($\gtrsim 4$ keV), noticeable fluctuations arise for large values of $\gamma_{\rm th}$. 
In this regime, the scattering is dominated by electrons with large Lorentz factors ($\gamma \gtrsim 40$), whose number density is strongly suppressed in the hybrid distribution. 
As a result, the effective number of sampled particles becomes small, leading increased statistical noise in the MC polarization estimates.

\subsection{Effects of the viewing angle on the IC Polarization}
\begin{figure}
\includegraphics[width=\columnwidth]{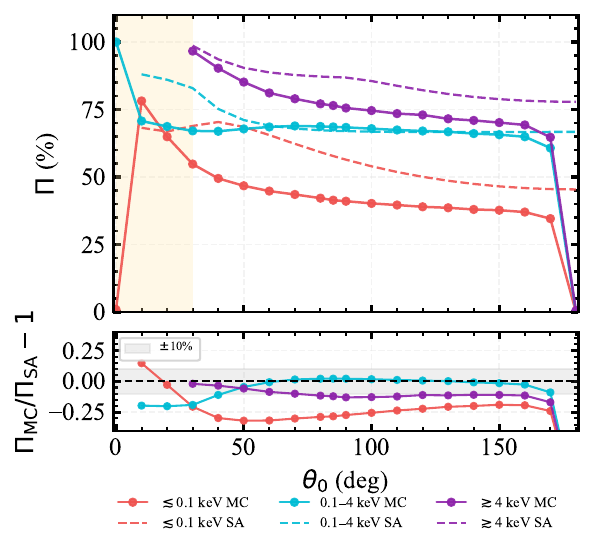}
\caption{IC polarization as a function of the viewing angle $\theta_{\rm o}$ for a hybrid electron distribution with $\gamma \in [1,100]$, $p_{\rm hybrid}=2$, and $\Theta=2$, scattering a monoenergetic, completely polarized photon beam at $\omega = 5\times10^{14}\,\mathrm{Hz}$. The degree of polarization $\Pi$ is shown for three energy bands: $\lesssim 0.1$ keV (red), $0.1\text{–}4$ keV (cyan), and $\gtrsim 4$ keV (purple). Solid symbols denote numerical results, while dashed curves show the corresponding SA predictions. The lower panel displays the relative difference between the numerical and SA results. Shaded regions mark viewing angles where photon statistics are insufficient to define the polarization reliably.}
\label{fig:fig8}
\end{figure}

We examine the dependence of the IC polarization on the viewing angle $\theta_{\rm o}$ for a fixed hybrid electron energy distribution, as shown in Figure~\ref{fig:fig8} for three representative energy bands.
For all viewing angles where the polarization is well defined,
we find that the polarization remains higher at higher photon energies and lower at lower energies. This demonstrates that the energy-dependent polarization behavior identified above is robust against variations in the viewing geometry. Changes in the viewing angle primarily affect the overall magnitude of the polarization, while the characteristic energy dependence associated with the electron distribution is preserved. The corresponding frequency-dependent intensity and polarization spectra for different viewing angles $\theta_{\rm o}$ are presented in 
Figure~\ref{fig:fig8_app_1},
illustrating how the spectral shape and polarization evolve with the line-of-sight orientation. 
The lower panel of Figure~\ref{fig:fig8} shows the relative differences between the numerical and SA results. 
The energy-dependent behavior of the deviations follows the same trends discussed in Figure~\ref{fig:fig7}. 
Discrepancies become more prominent at small scattering angles, where IC scattering into directions close to the incident photon direction is intrinsically less probable, resulting in limited photon statistics and increased numerical fluctuations. 
At $\theta_{\rm o} \approx 180^\circ$, the degree of polarization approaches zero because the scattering plane is not unique and all azimuthal orientations are symmetric, therefore the Stokes ($q/u$) contributions cancel out in the average. 

\section{Discussion and Conclusions}\label{Discussion and Conclusions}

In this work, we have investigated the polarization properties of inverse-Compton (IC) scattering produced by electrons following thermal, nonthermal power-law, and hybrid energy distributions. By combining numerical calculations with SA treatments, we systematically quantified how the underlying electron population shapes both the spectral and polarimetric behavior of the scattered radiation. Our main results can be summarized as follows:
\begin{itemize}
\item {\it Energy-dependent polarization as a diagnostic of hybrid electron populations} --
For hybrid electron energy distributions, the scattered emission exhibits a clear frequency-dependent structure that indicates the coexistence of a thermal distribution and a nonthermal tail. At low energies ($\lesssim 0.1$ keV), the emission is dominated by thermal electrons and displays polarization properties consistent with the purely thermal case. At high energies ($\gtrsim 4$ keV), the emission is governed primarily by the nonthermal component and approaches the polarization behavior characteristic of a pure power-law distribution. Between these limits, a transition band ($\sim 0.1\text{–}4$ keV) naturally emerges in which thermal and nonthermal contributions are comparable. The smooth evolution of the polarization degree across this band provides a direct and robust diagnostic of the hybrid nature of the electron population.
\item {\it Linear scaling to partially polarized seed photons} --
We show that, for hybrid electron distributions, the degree of polarization of the scattered radiation scales linearly with the polarization of the incident photons, extending the behavior previously established for purely nonthermal scattering to the hybrid case. When the incident radiation is only partially polarized, the IC process primarily rescales the polarization amplitude, while preserving the frequency-dependent polarization signatures imprinted by the electron energy distribution. This property significantly enhances the applicability of our model to realistic astrophysical environments.
\item {\it Expected polarization level in the X-ray band} --
In the observational X-ray band, our results indicate that inverse-Compton scattering by a hybrid electron population does not preserve a large polarization fraction when realistic seed photons are considered.

The scattered radiation is expected to exhibit only a few percent polarization when the incident photons originate from an accretion disk with low intrinsic polarization.
The relatively low observed polarization arises because the IC process acts as an intensity-weighted averaging of the polarization contributions. Although the scattering kernel is capable of producing a high degree of polarization for completely polarized incident radiation, the intrinsic polarization of disk photons is low. As a result, the final polarization level is significantly reduced.
\end{itemize}

\citet{Krawczynski2012ApJ} presented a MC formalism for IC polarization in the Thomson and Klein-Nishina regimes. Their Thomson-limit calculations were compared with SA results. In particular, the electron distributions considered in their study were primarily isotropic power-law distributions representing nonthermal electrons. 
In our work, we use the same MC framework of \citet{Krawczynski2012ApJ}. 
However, we use a smoothly joined hybrid electron energy distribution to examine the intrinsic polarization through the thermal-to-nonthermal transition. We focus on the polarization in the 0.2-4~keV band.

In accretion-powered systems, coronal geometry, disk reflection, and general relativistic photon propagation may also affect the observed polarization. 
\citet{Schnittman2010ApJ} used MC ray tracing in the Kerr metric to examine sandwich, clumpy, and spherical coronae. Their calculations showed that the degree of polarization depends on coronal structure and black-hole parameters. \citet{Beheshtipour2017ApJ} developed a general relativistic ray-tracing framework to study X-ray polarization from AGN coronae. 
Their X-ray polarization results were presented over the 0.1--100~keV band.
For high optical depths and hard spectra, their spherical-corona models gave higher degrees of polarization than their wedge-corona models in the 2-20 keV. 
These effects of coronal geometry and radiative transfer are complementary to the intrinsic electron-distribution effects considered here. To connect the intrinsic calculations to observed X-ray polarization, future models should include coronal geometries, disk reflection, and general relativistic photon propagation.
In addition to coronal geometry and radiative transfer effects, \citet{Beheshtipour2017ApJ} also examined the influence of the electron distribution on the polarization. They modeled the hybrid electron population as a thermal Maxwell-Jüttner component together with an additional, separate power-law component. In their test case, this power-law component accounted for 15\% of the total coronal electron energy, with $p=3$ and $\gamma\in[1,1000]$. They found that adding this separate nonthermal component produced little change in the observable polarization. A direct comparison of the predicted degrees of polarization between the results of \citet{Beheshtipour2017ApJ} and those in our work is not straightforward, because \citet{Beheshtipour2017ApJ} considered a different electron distribution and included additional radiative transfer effects. Their nonthermal component was a separate power-law distribution, whereas we adopt a smoothly joined hybrid distribution with $\Theta=2$, $p_{\rm hybrid}=2$ and $\gamma\in[1,100]$, with a smooth transition at $\gamma_{\rm th}$. We therefore focus on how the intrinsic polarization changes across the thermal-to-nonthermal transition produced by this electron distribution, particularly in the 0.2-4~keV band.

\citet{Moscibrodzka2022ApJS} developed a photon-conserving, covariant scattering kernel for thermal, power-law, hybrid, and $\kappa$ electron distributions. They used hybrid distributions as representative cases to validate the scattering kernel and the associated polarization calculations. 
Their calculations covered broad parameter ranges. 
The parameter $\gamma$ has the range of $1-10^6$. Two examples of electron temperature, $\Theta_{\rm e}=0.1$ and $100$, were given in their calculations. 
They presented radiation spectra over broad scattered energy ranges rather than a specific X-ray band. 
As a validation of our implementation, we reproduce the pure power-law test case presented in \citet{Moscibrodzka2022ApJS}. 
The main difference between \citet{Moscibrodzka2022ApJS} and our study is the scientific focus and the adopted electron distribution. In general, \citet{Moscibrodzka2022ApJS} primarily investigated the accuracy and applicability of the covariant scattering kernel using different electron distributions. In contrast, we study the intrinsic X-ray polarization produced by a smoothly joined hybrid electron distribution across the thermal-to-nonthermal transition. Specifically, we adopt $\Theta=2$ and $\gamma\in[1,100]$ to characterize the polarization in the 0.2-4~keV band. 
Therefore, our modeling results provide a reference for studies of the intrinsic polarization produced by IC scattering from hybrid electron energy distributions.

The electron temperatures considered in this work are expected to be common in astrophysical accretion environments. For example, in advection-dominated accretion flows (ADAFs), when the accretion rate approaches the critical value, the electron temperature in black-hole systems can reach $T_\text{e} \sim 10^{9\text{–}9.5}$ K at small radii. For accretion rates substantially below the critical threshold, the electron temperature can become even higher, exceeding $T_\text{e} \gtrsim 10^{10}$ K \citep{Narayan1995ApJ,yuan2014araa}. Such conditions naturally place the electron population in the relativistic regime. As a result, the polarization signatures derived here for relativistic thermal and hybrid electron distributions are possibly applicable to low-luminosity AGNs and black-hole XRBs in hard or quiescent states, where hot, optically thin accretion flows are expected to dominate the X-ray emission \citep{Esin1997ApJ,Done2007A&ARv}. Our results indicate that X-ray polarization measurements are sensitive to the underlying electron population. High-precision X-ray polarimetry missions, such as eXTP \citep{Zhang2025SCPMA}, are therefore expected to be capable of distinguishing the polarization contributions from thermal and nonthermal electrons.

\section*{Acknowledgements}
We appreciate the detailed suggestions made by the referee. This work has the financial support of the Natural Science Foundation of China 12393813, the National Key R\&D Program of China (2023YFE0101200), CSST grant CMS-CSST-2025-A07, and the Yunnan Revitalization Talent Support Program (YunLing Scholar Project). J.Y. is supported by the National Key R\&D Program of China (2021YFA1600402), the Strategic Priority Research Program of Chinese Academy of Sciences, grant No. XDB 41000000, the Natural Science Foundation of Yunnan Province (No. 202201AT070158 \& No. 202401AS070045), the Yunnan Revitalization Talent Support Program Young Talent Project, the National Natural Science Foundation of China (grants 12133011, 12288102), and the International Centre of Supernovae, Yunnan Key Laboratory (No. 202302AN360001). F.X is supported by National Key R\&D Program of China (grant No.2023YFE0117200), and National Natural Science Foundation of China (grant No.12373041 and No.12422306), and Bagui Scholars Program.

\textit{Software}: NumPy \citep{harris2020array}, Matplotlib \citep{Hunter2007}, TorchQuad \citep{Gomez2021JOSS}, and Scipy \citep{2020SciPy-NMeth} 

\section*{Data Availability}
The data underlying this article are available in the article and in the cited references.

\bibliographystyle{mnras}
\bibliography{references}

\begin{thebibliography}{}
\makeatletter
\relax
\def\mn@urlcharsother{\let\do\@makeother \do\$\do\&\do\#\do\^\do\_\do\%\do\~}
\def\mn@doi{\begingroup\mn@urlcharsother \@ifnextchar [ {\mn@doi@}
  {\mn@doi@[]}}
\def\mn@doi@[#1]#2{\def\@tempa{#1}\ifx\@tempa\@empty \href
  {http://dx.doi.org/#2} {doi:#2}\else \href {http://dx.doi.org/#2} {#1}\fi
  \endgroup}
\def\mn@eprint#1#2{\mn@eprint@#1:#2::\@nil}
\def\mn@eprint@arXiv#1{\href {http://arxiv.org/abs/#1} {{\tt arXiv:#1}}}
\def\mn@eprint@dblp#1{\href {http://dblp.uni-trier.de/rec/bibtex/#1.xml}
  {dblp:#1}}
\def\mn@eprint@#1:#2:#3:#4\@nil{\def\@tempa {#1}\def\@tempb {#2}\def\@tempc
  {#3}\ifx \@tempc \@empty \let \@tempc \@tempb \let \@tempb \@tempa \fi \ifx
  \@tempb \@empty \def\@tempb {arXiv}\fi \@ifundefined
  {mn@eprint@\@tempb}{\@tempb:\@tempc}{\expandafter \expandafter \csname
  mn@eprint@\@tempb\endcsname \expandafter{\@tempc}}}

\bibitem[\protect\citeauthoryear{{Beheshtipour}, {Krawczynski}  \&
  {Malzac}}{{Beheshtipour} et~al.}{2017}]{Beheshtipour2017ApJ}
{Beheshtipour} B.,  {Krawczynski} H.,   {Malzac} J.,  2017, \mn@doi [\apj]
  {10.3847/1538-4357/aa906a}, \href
  {https://ui.adsabs.harvard.edu/abs/2017ApJ...850...14B} {850, 14}

\bibitem[\protect\citeauthoryear{{Bonometto}, {Cazzola}  \&
  {Saggion}}{{Bonometto} et~al.}{1970}]{Bonometto1970A&A}
{Bonometto} S.,  {Cazzola} P.,   {Saggion} A.,  1970, \aap, \href
  {https://ui.adsabs.harvard.edu/abs/1970A&A.....7..292B} {7, 292}

\bibitem[\protect\citeauthoryear{{Canfield}, {Howard}  \& {Liang}}{{Canfield}
  et~al.}{1987}]{Canfield1987ApJ}
{Canfield} E.,  {Howard} W.~M.,   {Liang} E.~P.,  1987, \mn@doi [\apj]
  {10.1086/165853}, \href
  {https://ui.adsabs.harvard.edu/abs/1987ApJ...323..565C} {323, 565}

\bibitem[\protect\citeauthoryear{{Chang}, {Jiang}  \& {Lin}}{{Chang}
  et~al.}{2014}]{2014Chang_apj}
{Chang} Z.,  {Jiang} Y.,   {Lin} H.-N.,  2014, \mn@doi [\apj]
  {10.1088/0004-637X/780/1/68}, \href
  {https://ui.adsabs.harvard.edu/abs/2014ApJ...780...68C} {780, 68}

\bibitem[\protect\citeauthoryear{{Compton}}{{Compton}}{1923}]{Compton1923PhRv}
{Compton} A.~H.,  1923, \mn@doi [Physical Review] {10.1103/PhysRev.21.483},
  \href {https://ui.adsabs.harvard.edu/abs/1923PhRv...21..483C} {21, 483}

\bibitem[\protect\citeauthoryear{{Coppi}}{{Coppi}}{1999}]{Coppi1999ASPC}
{Coppi} P.~S.,  1999, in {Poutanen} J.,  {Svensson} R.,  eds,  Astronomical
  Society of the Pacific Conference Series Vol. 161, High Energy Processes in
  Accreting Black Holes. p.~375 (\mn@eprint {arXiv} {astro-ph/9903158}),
  \mn@doi{10.48550/arXiv.astro-ph/9903158}

\bibitem[\protect\citeauthoryear{{Davelaar}, {Ryan}, {Wong}, {Bronzwaer},
  {Olivares}, {Mo{\'s}cibrodzka}, {Gammie}  \& {Falcke}}{{Davelaar}
  et~al.}{2023}]{Davelaar2023MNRAS}
{Davelaar} J.,  {Ryan} B.~R.,  {Wong} G.~N.,  {Bronzwaer} T.,  {Olivares} H.,
  {Mo{\'s}cibrodzka} M.,  {Gammie} C.~F.,   {Falcke} H.,  2023, \mn@doi
  [\mnras] {10.1093/mnras/stad3023}, \href
  {https://ui.adsabs.harvard.edu/abs/2023MNRAS.526.5326D} {526, 5326}

\bibitem[\protect\citeauthoryear{{De Young}}{{De Young}}{1966}]{Young1966JMP}
{De Young} D.~S.,  1966, \mn@doi [Journal of Mathematical Physics]
  {10.1063/1.1704844}, \href
  {https://ui.adsabs.harvard.edu/abs/1966JMP.....7.1916D} {7, 1916}

\bibitem[\protect\citeauthoryear{Dolence, Gammie, Mościbrodzka  \&
  Leung}{Dolence et~al.}{2009}]{Dolence_2009}
Dolence J.~C.,  Gammie C.~F.,  Mościbrodzka M.,   Leung P.~K.,  2009, \mn@doi
  [\apjs] {10.1088/0067-0049/184/2/387}, 184, 387

\bibitem[\protect\citeauthoryear{{Done}, {Gierli{\'n}ski}  \& {Kubota}}{{Done}
  et~al.}{2007}]{Done2007A&ARv}
{Done} C.,  {Gierli{\'n}ski} M.,   {Kubota} A.,  2007, \mn@doi [\aapr]
  {10.1007/s00159-007-0006-1}, \href
  {https://ui.adsabs.harvard.edu/abs/2007A&ARv..15....1D} {15, 1}

\bibitem[\protect\citeauthoryear{{Dreyer} \& {B{\"o}ttcher}}{{Dreyer} \&
  {B{\"o}ttcher}}{2021}]{Dreyer2021ApJ}
{Dreyer} L.,  {B{\"o}ttcher} M.,  2021, \mn@doi [\apj]
  {10.3847/1538-4357/abc9b8}, \href
  {https://ui.adsabs.harvard.edu/abs/2021ApJ...906...18D} {906, 18}

\bibitem[\protect\citeauthoryear{{Eardley}, {Lightman}  \& {Shapiro}}{{Eardley}
  et~al.}{1975}]{Eardley1975ApJ}
{Eardley} D.~M.,  {Lightman} A.~P.,   {Shapiro} S.~L.,  1975, \mn@doi [\apjl]
  {10.1086/181871}, \href
  {https://ui.adsabs.harvard.edu/abs/1975ApJ...199L.153E} {199, L153}

\bibitem[\protect\citeauthoryear{{Esin}, {McClintock}  \& {Narayan}}{{Esin}
  et~al.}{1997}]{Esin1997ApJ}
{Esin} A.~A.,  {McClintock} J.~E.,   {Narayan} R.,  1997, \mn@doi [\apj]
  {10.1086/304829}, \href
  {https://ui.adsabs.harvard.edu/abs/1997ApJ...489..865E} {489, 865}

\bibitem[\protect\citeauthoryear{{Ewing} et~al.,}{{Ewing}
  et~al.}{2025}]{Ewing2025MNRAS}
{Ewing} M.,  et~al., 2025, \mn@doi [\mnras] {10.1093/mnras/staf859}, \href
  {https://ui.adsabs.harvard.edu/abs/2025MNRAS.tmp..846E} {541, 1774}

\bibitem[\protect\citeauthoryear{{Fabian}, {Lohfink}, {Belmont}, {Malzac}  \&
  {Coppi}}{{Fabian} et~al.}{2017}]{Fabian2017MNRAS}
{Fabian} A.~C.,  {Lohfink} A.,  {Belmont} R.,  {Malzac} J.,   {Coppi} P.,
  2017, \mn@doi [\mnras] {10.1093/mnras/stx221}, \href
  {https://ui.adsabs.harvard.edu/abs/2017MNRAS.467.2566F} {467, 2566}

\bibitem[\protect\citeauthoryear{{Fano}}{{Fano}}{1949}]{Fano1949JOSA}
{Fano} U.,  1949, \mn@doi [JOSA] {10.1364/JOSA.39.000859}, \href
  {https://ui.adsabs.harvard.edu/abs/1949JOSA...39..859F} {39, 859}

\bibitem[\protect\citeauthoryear{{Fano}}{{Fano}}{1957}]{Fano1957RvMP}
{Fano} U.,  1957, \mn@doi [RvMP] {10.1103/RevModPhys.29.74}, \href
  {https://ui.adsabs.harvard.edu/abs/1957RvMP...29...74F} {29, 74}

\bibitem[\protect\citeauthoryear{{Feng} \& {Bellazzini}}{{Feng} \&
  {Bellazzini}}{2020}]{Feng2020NatAs}
{Feng} H.,  {Bellazzini} R.,  2020, \mn@doi [Nature Astronomy]
  {10.1038/s41550-020-1103-6}, \href
  {https://ui.adsabs.harvard.edu/abs/2020NatAs...4..547F} {4, 547}

\bibitem[\protect\citeauthoryear{{Giannios} \& {Spitkovsky}}{{Giannios} \&
  {Spitkovsky}}{2009}]{2009Giannios_mnras}
{Giannios} D.,  {Spitkovsky} A.,  2009, \mn@doi [\mnras]
  {10.1111/j.1365-2966.2009.15454.x}, \href
  {https://ui.adsabs.harvard.edu/abs/2009MNRAS.400..330G} {400, 330}

\bibitem[\protect\citeauthoryear{{G{\'o}mez}, {Toftevaag}  \&
  {Meoni}}{{G{\'o}mez} et~al.}{2021}]{Gomez2021JOSS}
{G{\'o}mez} P.,  {Toftevaag} H.,   {Meoni} G.,  2021, \mn@doi [The Journal of
  Open Source Software] {10.21105/joss.03439}, \href
  {https://ui.adsabs.harvard.edu/abs/2021JOSS....6.3439G} {6, 3439}

\bibitem[\protect\citeauthoryear{{Haardt} \& {Maraschi}}{{Haardt} \&
  {Maraschi}}{1991}]{Haardt1991ApJ}
{Haardt} F.,  {Maraschi} L.,  1991, \mn@doi [\apjl] {10.1086/186171}, \href
  {https://ui.adsabs.harvard.edu/abs/1991ApJ...380L..51H} {380, L51}

\bibitem[\protect\citeauthoryear{{Haardt} \& {Maraschi}}{{Haardt} \&
  {Maraschi}}{1993}]{Haardt1993ApJ}
{Haardt} F.,  {Maraschi} L.,  1993, \mn@doi [\apj] {10.1086/173020}, \href
  {https://ui.adsabs.harvard.edu/abs/1993ApJ...413..507H} {413, 507}

\bibitem[\protect\citeauthoryear{{Haardt} \& {Matt}}{{Haardt} \&
  {Matt}}{1993}]{Haardt1993MNRAS}
{Haardt} F.,  {Matt} G.,  1993, \mn@doi [\mnras] {10.1093/mnras/261.2.346},
  \href {https://ui.adsabs.harvard.edu/abs/1993MNRAS.261..346H} {261, 346}

\bibitem[\protect\citeauthoryear{Harris et~al.,}{Harris
  et~al.}{2020}]{harris2020array}
Harris C.~R.,  et~al., 2020, \mn@doi [Nature] {10.1038/s41586-020-2649-2}, 585,
  357

\bibitem[\protect\citeauthoryear{Hunter}{Hunter}{2007}]{Hunter2007}
Hunter J.~D.,  2007, \mn@doi [Computing in Science \& Engineering]
  {10.1109/MCSE.2007.55}, 9, 90

\bibitem[\protect\citeauthoryear{{Ingram} et~al.,}{{Ingram}
  et~al.}{2023}]{2023Ingram}
{Ingram} A.,  et~al., 2023, \mn@doi [\mnras] {10.1093/mnras/stad2625}, \href
  {https://ui.adsabs.harvard.edu/abs/2023MNRAS.525.5437I} {525, 5437}

\bibitem[\protect\citeauthoryear{{Kochanek}}{{Kochanek}}{2004}]{Kochanek2004ApJ}
{Kochanek} C.~S.,  2004, \mn@doi [\apj] {10.1086/382180}, \href
  {https://ui.adsabs.harvard.edu/abs/2004ApJ...605...58K} {605, 58}

\bibitem[\protect\citeauthoryear{{Krawczynski}}{{Krawczynski}}{2012}]{Krawczynski2012ApJ}
{Krawczynski} H.,  2012, \mn@doi [\apj] {10.1088/0004-637X/744/1/30}, \href
  {https://ui.adsabs.harvard.edu/abs/2012ApJ...744...30K} {744, 30}

\bibitem[\protect\citeauthoryear{{Krawczynski} et~al.,}{{Krawczynski}
  et~al.}{2022}]{Krawczynski2022Sci}
{Krawczynski} H.,  et~al., 2022, \mn@doi [Science] {10.1126/science.add5399},
  \href {https://ui.adsabs.harvard.edu/abs/2022Sci...378..650K} {378, 650}

\bibitem[\protect\citeauthoryear{{Kushwaha}, {Jayasurya}, {Agrawal}  \&
  {Nandi}}{{Kushwaha} et~al.}{2023}]{Kushwaha2023MNRAS}
{Kushwaha} A.,  {Jayasurya} K.~M.,  {Agrawal} V.~K.,   {Nandi} A.,  2023,
  \mn@doi [\mnras] {10.1093/mnrasl/slad070}, \href
  {https://ui.adsabs.harvard.edu/abs/2023MNRAS.524L..15K} {524, L15}

\bibitem[\protect\citeauthoryear{{Lightman} \& {Shapiro}}{{Lightman} \&
  {Shapiro}}{1976}]{Lightman1976ApJ}
{Lightman} A.~P.,  {Shapiro} S.~L.,  1976, \mn@doi [\apj] {10.1086/154131},
  \href {https://ui.adsabs.harvard.edu/abs/1976ApJ...203..701L} {203, 701}

\bibitem[\protect\citeauthoryear{{Liu}, {Mao}  \& {Liu}}{{Liu}
  et~al.}{2024}]{Liu2024MNRAS}
{Liu} J.-Y.,  {Mao} J.,   {Liu} B.~F.,  2024, \mn@doi [\mnras]
  {10.1093/mnras/stad3615}, \href
  {https://ui.adsabs.harvard.edu/abs/2024MNRAS.527.5627L} {527, 5627}

\bibitem[\protect\citeauthoryear{{Marinucci} et~al.,}{{Marinucci}
  et~al.}{2022}]{Marinucci2022MNRAS}
{Marinucci} A.,  et~al., 2022, \mn@doi [\mnras] {10.1093/mnras/stac2634}, \href
  {https://ui.adsabs.harvard.edu/abs/2022MNRAS.516.5907M} {516, 5907}

\bibitem[\protect\citeauthoryear{Matt, Feroci, Rapisarda  \& Costa}{Matt
  et~al.}{1996}]{matt1996}
Matt G.,  Feroci M.,  Rapisarda M.,   Costa E.,  1996, \mn@doi [RaPC]
  {10.1016/0969-806X(95)00472-A}, 48, 403

\bibitem[\protect\citeauthoryear{{McMaster}}{{McMaster}}{1961}]{McMaster1961RvMP}
{McMaster} W.~H.,  1961, \mn@doi [Reviews of Modern Physics]
  {10.1103/RevModPhys.33.8}, \href
  {https://ui.adsabs.harvard.edu/abs/1961RvMP...33....8M} {33, 8}

\bibitem[\protect\citeauthoryear{{Mo{\'s}cibrodzka}}{{Mo{\'s}cibrodzka}}{2022}]{Moscibrodzka2022ApJS}
{Mo{\'s}cibrodzka} M.~A.,  2022, \mn@doi [\apjs] {10.3847/1538-4365/ac972c},
  \href {https://ui.adsabs.harvard.edu/abs/2022ApJS..263....6M} {263, 6}

\bibitem[\protect\citeauthoryear{Nagirner \& Poutanen}{Nagirner \&
  Poutanen}{1993}]{nagirner1993}
Nagirner D.~I.,  Poutanen J.,  1993, \aap, \href
  {https://ui.adsabs.harvard.edu/abs/1993A&A...275..325N} {275, 325}

\bibitem[\protect\citeauthoryear{{Narayan} \& {Yi}}{{Narayan} \&
  {Yi}}{1995}]{Narayan1995ApJ}
{Narayan} R.,  {Yi} I.,  1995, \mn@doi [\apj] {10.1086/176343}, \href
  {https://ui.adsabs.harvard.edu/abs/1995ApJ...452..710N} {452, 710}

\bibitem[\protect\citeauthoryear{{Pandya}, {Zhang}, {Chandra}  \&
  {Gammie}}{{Pandya} et~al.}{2016}]{2016ApJ...822...34P}
{Pandya} A.,  {Zhang} Z.,  {Chandra} M.,   {Gammie} C.~F.,  2016, \mn@doi
  [\apj] {10.3847/0004-637X/822/1/34}, \href
  {https://ui.adsabs.harvard.edu/abs/2016ApJ...822...34P} {822, 34}

\bibitem[\protect\citeauthoryear{Poutanen \& Vilhu}{Poutanen \&
  Vilhu}{1993}]{poutanen1993A&A}
Poutanen J.,  Vilhu O.,  1993, \aap, 275, 337

\bibitem[\protect\citeauthoryear{{Ratheesh} et~al.,}{{Ratheesh}
  et~al.}{2024}]{Ratheesh2024ApJ}
{Ratheesh} A.,  et~al., 2024, \mn@doi [\apj] {10.3847/1538-4357/ad226e}, \href
  {https://ui.adsabs.harvard.edu/abs/2024ApJ...964...77R} {964, 77}

\bibitem[\protect\citeauthoryear{Rawat, Garg  \& M{\'e}ndez}{Rawat
  et~al.}{2023}]{rawat2023detection}
Rawat D.,  Garg A.,   M{\'e}ndez M.,  2023, The Astrophysical Journal Letters,
  949, L43

\bibitem[\protect\citeauthoryear{{Rees}}{{Rees}}{1975}]{Rees1975MNRAS}
{Rees} M.~J.,  1975, \mn@doi [\mnras] {10.1093/mnras/171.3.457}, \href
  {https://ui.adsabs.harvard.edu/abs/1975MNRAS.171..457R} {171, 457}

\bibitem[\protect\citeauthoryear{{Reis} \& {Miller}}{{Reis} \&
  {Miller}}{2013}]{Reis2013ApJ}
{Reis} R.~C.,  {Miller} J.~M.,  2013, \mn@doi [\apjl]
  {10.1088/2041-8205/769/1/L7}, \href
  {https://ui.adsabs.harvard.edu/abs/2013ApJ...769L...7R} {769, L7}

\bibitem[\protect\citeauthoryear{{Schnittman} \& {Krolik}}{{Schnittman} \&
  {Krolik}}{2010}]{Schnittman2010ApJ}
{Schnittman} J.~D.,  {Krolik} J.~H.,  2010, \mn@doi [\apj]
  {10.1088/0004-637X/712/2/908}, \href
  {https://ui.adsabs.harvard.edu/abs/2010ApJ...712..908S} {712, 908}

\bibitem[\protect\citeauthoryear{{Thorne} \& {Price}}{{Thorne} \&
  {Price}}{1975}]{Thorne1975ApJ}
{Thorne} K.~S.,  {Price} R.~H.,  1975, \mn@doi [\apjl] {10.1086/181720}, \href
  {https://ui.adsabs.harvard.edu/abs/1975ApJ...195L.101T} {195, L101}

\bibitem[\protect\citeauthoryear{{Ursini}, {Matt}, {Bianchi}, {Marinucci},
  {Dov{\v{c}}iak}  \& {Zhang}}{{Ursini} et~al.}{2022}]{Ursini2022MNRAS}
{Ursini} F.,  {Matt} G.,  {Bianchi} S.,  {Marinucci} A.,  {Dov{\v{c}}iak} M.,
  {Zhang} W.,  2022, \mn@doi [\mnras] {10.1093/mnras/stab3745}, \href
  {https://ui.adsabs.harvard.edu/abs/2022MNRAS.510.3674U} {510, 3674}

\bibitem[\protect\citeauthoryear{{Veledina} et~al.,}{{Veledina}
  et~al.}{2023}]{Veledina2023ApJ}
{Veledina} A.,  et~al., 2023, \mn@doi [\apjl] {10.3847/2041-8213/ad0781}, \href
  {https://ui.adsabs.harvard.edu/abs/2023ApJ...958L..16V} {958, L16}

\bibitem[\protect\citeauthoryear{Virtanen et~al.,}{Virtanen
  et~al.}{2020}]{2020SciPy-NMeth}
Virtanen P.,  et~al., 2020, \mn@doi [Nature Methods]
  {10.1038/s41592-019-0686-2}, \href {https://rdcu.be/b08Wh} {17, 261}

\bibitem[\protect\citeauthoryear{{Weisskopf} et~al.,}{{Weisskopf}
  et~al.}{2022}]{2022Weisskopf}
{Weisskopf} M.~C.,  et~al., 2022, \mn@doi [JATIS] {10.1117/1.JATIS.8.2.026002},
  \href {https://ui.adsabs.harvard.edu/abs/2022JATIS...8b6002W} {8, 026002}

\bibitem[\protect\citeauthoryear{{Yang}, {Wang}, {Yang}  \& {Yuan}}{{Yang}
  et~al.}{2021}]{Yang2021ApJS}
{Yang} X.-L.,  {Wang} J.-C.,  {Yang} C.-Y.,   {Yuan} Z.-L.,  2021, \mn@doi
  [\apjs] {10.3847/1538-4365/abec73}, \href
  {https://ui.adsabs.harvard.edu/abs/2021ApJS..254...29X} {254, 29}

\bibitem[\protect\citeauthoryear{Yuan \& Narayan}{Yuan \&
  Narayan}{2014}]{yuan2014araa}
Yuan F.,  Narayan R.,  2014, \mn@doi [Annual Review of Astronomy and
  Astrophysics] {10.1146/annurev-astro-082812-141003}, 52, 529

\bibitem[\protect\citeauthoryear{{Zhang} et~al.,}{{Zhang}
  et~al.}{2025}]{Zhang2025SCPMA}
{Zhang} S.-N.,  et~al., 2025, \mn@doi [Science China Physics, Mechanics, and
  Astronomy] {10.1007/s11433-025-2786-6}, \href
  {https://ui.adsabs.harvard.edu/abs/2025SCPMA..6819502Z} {68, 119502}

\makeatother
\end{thebibliography}

\appendix
\section{Monte Carlo Simulation of the Polarization in IC Scattering}\label{appendixa}
In this work, we simulate the polarization of IC emission using a MC approach. The present approach is based on the framework developed by \citet{Krawczynski2012ApJ}. The numerical method is summarized below. To illustrate the general process of IC polarization, we repeat some important formulas presented by \citet{Krawczynski2012ApJ}.
All calculations start and end in the plasma frame (PF). The reference coordinate system in the PF is established using a right-handed set of unit basis vectors, denoted as $C\equiv\{{\bf \hat{x},\hat{y},\hat{z}} \}$. We examine the interaction between a collimated, monoenergetic photon beam and an electron population of identical energy, where both species are initially unidirectional. The incident photons travel along the $\hat{\mathbf{z}}$-axis, with four-momentum given by $k_{\mathrm{i}} = (\omega_{\mathrm{i}}/c,\, \mathbf{k}_{\mathrm{i}})$, and direction $\hat{\mathbf{k}}_{\mathrm{i}} = \mathbf{k}_{\mathrm{i}}/|\mathbf{k}_{\mathrm{i}}|$. The electrons possess four-momentum $p = (\gamma m_e c,\, \mathbf{p})$, and their propagation direction is denoted by $\hat{\mathbf{p}} = \mathbf{p}/|\mathbf{p}|$.
We utilize the dimensionless Stokes vector $\mathbf{s}=(i,q,u)$ to describe the simulated photons, where the component $i$ represents the statistical intensity, and the ratios $q/i$ and $u/i$ define the degree of linear polarization. This formulation is based on the quantum-mechanical framework for Stokes parameters as outlined by \citet{McMaster1961RvMP}. We define the Stokes vector $\mathbf{s}_{\rm a}$ with respect to a local coordinate system $C_{\rm a}\,\equiv\{\mathbf{\hat{x}}_{\rm a},\mathbf{\hat{y}}_{\rm a}, \mathbf{\hat{z}}_{\rm a}\}$, where the $\mathbf{\hat{z}}_{\rm a}$-axis is consistently set to align with the photon's propagation direction. For convenience, when specifying a new system, only the $\mathbf{\hat{y}}_{\rm a}$ and $\mathbf{\hat{z}}_{\rm a}$ basis vectors will be provided, as the $\mathbf{\hat{x}}_{\rm a}$ vector is readily determined by the right-hand rule $\mathbf{\hat{x}}_{\rm a}=\mathbf{\hat{y}}_{\rm a}\times\mathbf{\hat{z}}_{\rm a}$. We adopt the standard convention for linear polarization defined within the local frame $C_{\rm a}$: a $100\%$ linearly polarized beam with the electric field $\hat{\boldsymbol e}$ parallel to the $\mathbf{\hat{y}}_{\rm a}$ axis is denoted by the Stokes vector $\mathbf{s}_{\rm a}=(1,1,0)$ (i.e., $\mathbf{\hat{q}}_+ \equiv \mathbf{\hat{y}}_{\rm a}$). Conversely, polarization parallel to the $\mathbf{\hat{x}}_{\rm a}$ axis corresponds to $\mathbf{s}_{\rm a}=(1,-1,0)$ (i.e., $\mathbf{\hat{q}}_- \equiv \mathbf{\hat{x}}_{\rm a}$). Furthermore, the vectors $\mathbf{s}_{\rm a}=(1,0,1)$ and $\mathbf{s}_{\rm a}=(1,0,-1)$ represent polarization along the $\mathbf{\hat{u}}_+$ and $\mathbf{\hat{u}}_-$ directions, where the diagonal unit vectors are defined as $\mathbf{\hat{u}}_+ \equiv (\mathbf{\hat{y}}_{\rm a}-\mathbf{\hat{x}}_{\rm a})/\sqrt{2}$ and $\mathbf{\hat{u}}_- \equiv (\mathbf{\hat{y}}_{\rm a}+\mathbf{\hat{x}}_{\rm a})/\sqrt{2}$, respectively. The degree of polarization $\Pi$ is given by $\Pi=\sqrt{q^2+u^2}/i$.
When transforming the Stokes vector $\mathbf{s}_0$ from the coordinate system $C_0$ to $C_1$, the resulting vector $\mathbf{s}_1$ is obtained via a matrix operation: $\mathbf{s}_1 = \mathbf{M}[\chi]\mathbf{s}_0$. This transformation applies when $C_1$ is related to $C_0$ by a counterclockwise rotation (viewed along $-\mathbf{\hat{z}}_{\rm i}$) of an angle $\chi$ around the common $z$-axis. The rotation matrix $\mathbf{M}[\chi]$ is defined as
$$\mathbf{M}[\chi]=
\begin{pmatrix}
1 & 0 & 0 \\
0 & \cos 2\chi & \sin 2\chi \\
0 & -\sin 2\chi & \cos 2\chi
\end{pmatrix}.$$
For convenience, we introduce three auxiliary functions. First, the normalization function $\hat{n}[\mathbf{v}]$ returns the unit vector of its argument, defined as $\hat{n}[\mathbf{v}] = \mathbf{v}/\sqrt{\mathbf{v} \cdot \mathbf{v}}$.
Second, the normalized component of the vector $\mathbf{v}$ that is perpendicular to the unit vector $\hat{\mathbf{w}}$ is computed using the function $\hat{P}[\mathbf{v},\mathbf{w}]$:
$\hat{P}[\mathbf{v},\mathbf{w}] = \hat{n}[\mathbf{v} - (\mathbf{v} \cdot \hat{\mathbf{w}})\hat{\mathbf{w}}]$. Finally, the function $\Delta[C_A, C_B]$ determines the rotation angle between two coordinate systems, $C_A$ and $C_B$, that share a common $z$-axis: $\Delta[C_A, C_B] = -\mathrm{sign}(\hat{\mathbf{x}}_A \cdot \hat{\mathbf{y}}_B)\, \arccos(\hat{\mathbf{y}}_A \cdot \hat{\mathbf{y}}_B).$
A positive result for $\Delta$ indicates that $C_B$ is obtained by a counterclockwise rotation of $C_A$ around the $\hat{\mathbf{z}}_A$-axis, where $\mathrm{sign}(u)$ equals $+1$ if $u>0$, $-1$ if $u<0$, and $0 $ if $u=0$.

First, the simulation commences by randomly drawing the direction $(\theta_{\rm e}, \phi_{\rm e})$ of the target electron, represented by the unit vector $\hat{\mathbf{p}}$, within the plasma frame (PF). The probability of an electron-photon interaction is simultaneously tracked by weighting the incident photon's Stokes vector $\mathbf{s}$ with the factor $(1-\beta \mu_{k_{\rm i}p})$. Here, $\beta$ is the electron velocity normalized by the speed of light, and $\mu_{k_{\rm i}p}=\hat{\mathbf{k}}_{\rm i} \cdot \hat{\mathbf{p}}$ is the cosine of the angle between the incident photon direction $\hat{\mathbf{k}}_{\rm i}$ and the electron direction.

Second, the incident photon's wave vector $\mathbf{k}_{\rm i}$ and Stokes vector $\mathbf{s}$ are Lorentz transformed into the Electron Rest Frame (ERF) using the following relations. The cosine of the angle between the photon's propagation direction $\hat{\mathbf{k}}_{\rm i}'$ and the electron's direction $\hat{\mathbf{p}}'$ in the ERF, $\mu_{k_{\rm i}'p'}'$, is given by

$$\mu_{k_{\rm i}'p'}'\,=\,\frac{\mu_{k_{\rm i}p}-\beta}{1-\mu_{k_{\rm i}p}\beta}$$

The transformed wave vector $\hat{\mathbf{k}}_{\rm i}'$ is then determined by

$$\hat{\mathbf{k}}_{\rm i}'\,=\,\mu_{k_{\rm i}'p'}' \hat{\mathbf{p}}+\sqrt{1-\mu_{k_{\rm i}'p'}'{}^2} \,\hat{P}[\hat{\mathbf{k}}_{\rm i},\hat{\mathbf{p}}].$$

The photon frequency $\omega$ is transformed to $\omega'$ according to $\omega'=\gamma (1-\beta \mu_{k_{\rm i}p}) \omega$, where $\gamma=\sqrt{1/(1-\beta^2)}$ is the electron's Lorentz factor.
Following \citet{Young1966JMP}, the degree of linear polarization $\Pi$ is invariant under Lorentz boosts. In our application, where the Stokes vector $\mathbf{s}$ tracks the statistical weight and polarization direction but not the beam intensity, the component $i$ is consequently unaffected by the transformation. Consequently, to facilitate the subsequent scattering calculation, the Stokes vector $\mathbf{s}$ is first rotated into a coordinate system $C_{k_{\rm i}p}$ where the $y$-axis is perpendicular to the $\hat{\mathbf{k}}_{\rm i}$-$\hat{\mathbf{p}}$ plane. This rotation yields $\mathbf{s}_{k_{\rm i}p} = \mathbf{M}[\phi_{\rm e}] \mathbf{s}$, where $C_{k_{\rm i}p} = \{\hat{\mathbf{n}}[\hat{\mathbf{k}}_{\rm i} \times \hat{\mathbf{p}}], \hat{\mathbf{k}}_{\rm i}\}$. 
Crucially, the resulting Stokes vector $\mathbf{s}_{k_{\rm i}p}$ in the PF is invariant under the Lorentz boost when compared to the Stokes vector $\mathbf{s}_{k_{\rm i}'p}'$ defined in the ERF relative to the corresponding transformed system, $C_{k_{\rm i}'p}'$ $$\mathbf{s}_{k_{\rm i}'p}'\,=\,\mathbf{s}_{k_{\rm i}p}.$$Here, $C_{k_{\rm i}'p}'$ is defined relative to the incident wave vector in the ERF $C_{k_{\rm i}'p}' = \{\hat{\mathbf{n}}[\hat{\mathbf{k}}_{\rm i}' \times \hat{\mathbf{p}}], \hat{\mathbf{k}}_{\rm i}'\}$.
The third step simulates the Compton scattering event entirely within the ERF.
We begin by randomly selecting the direction of the scattered photon, denoted by the unit vector $\hat{\mathbf{k}}_{\rm o}'$, by sampling uniformly over solid angle for numerical convenience. This sampling does not imply isotropic scattering. The scattering angle is $\theta_{\rm o}' = \arccos(\mu_{k_{\rm i}'k_{\rm o}'}')$, where $\mu_{k_{\rm i}'k_{\rm o}'}' = \hat{\mathbf{k}}_{\rm i}' \cdot \hat{\mathbf{k}}_{\rm o}'$. The frequency of the scattered photon, $\omega_{\rm o}'$, is then determined using the Compton scattering formula \citep{Compton1923PhRv} $$\omega_{\rm o}'\,=\,\frac{1}{1+\epsilon'\,(1-\mu_{k_{\rm i}'k_{\rm o}'}')}\omega'$$where $\omega'$ is the incident frequency and $\epsilon'=\hbar \omega'/m_{\rm e}c^2$ is the target photon energy in units of the electron rest mass. The polarization state of the scattered photon, $\mathbf{s}_{k_{\rm o}'k_{\rm i}'}'$, is subsequently calculated by applying Fano's scattering matrix $\mathbf{F}$ to the incident Stokes vector $\mathbf{s}_{k_{\rm i}'p}'$ \citep{Fano1949JOSA, Fano1957RvMP} $$\mathbf{s}_{k_{\rm o}'k_{\rm i}'}'\,=\,\mathbf{F}[\theta_{\rm s}',\epsilon',\epsilon_{\rm o}']\, \mathbf{M}[\Delta[C_{k_{\rm i}'p},C_{k_{\rm o}'k_{\rm i}'}]]\, \mathbf{s}_{k_{\rm i}'p}'.$$Here, $\mathbf{M}[\Delta]$ is the rotation matrix that transforms $\mathbf{s}_{k_{\rm i}'p}'$ into the coordinate frame $C_{k_{\rm o}'k_{\rm i}'} = \{\hat{\mathbf{n}}[\hat{\mathbf{k}}_{\rm o}' \times \hat{\mathbf{k}}_{\rm i}'], \hat{\mathbf{k}}_{\rm o}'\}$ whose $y$-axis is perpendicular to the scattering plane ($\hat{\mathbf{k}}_{\rm i}'$-$\hat{\mathbf{k}}_{\rm o}'$ plane)—a necessary prerequisite for applying Fano's matrix. Fano's matrix $\mathbf{F}$ (which accounts for both Thomson and Klein-Nishina regimes and encodes the angular dependence of the differential cross-section. This effectively assigns the correct statistical weight to each scattering event) is given by 
\begin{equation*}
\setlength{\arraycolsep}{1.0pt}
\small
\begin{aligned}
\mathbf{F}(\theta_{\rm o}',\epsilon',\epsilon_{\rm o}')
={}&
\left(\frac{\epsilon_{\rm o}'}{\epsilon'}\right)^2 \\
&\quad \times
\begin{pmatrix}
1+\cos^2\theta_{\rm o}'
+(\epsilon'-\epsilon_{\rm o}')(1-\cos\theta_{\rm o}')
& \sin^2\theta_{\rm o}' & 0 \\
\sin^2\theta_{\rm o}'
& 1+\cos^2\theta_{\rm o}' & 0 \\
0 & 0 & 2\cos\theta_{\rm o}'
\end{pmatrix}.
\end{aligned}
\end{equation*}
where $\epsilon_{\rm o}'=\hbar \omega_{\rm o}'/m_{\rm e}c^2$ is the scattered photon energy normalized to the electron rest mass. For simplicity, we omit the numerical factor $r_0^2/2$ (classical electron radius $r_0$) from $\mathbf{F}$ as we do not track absolute fluxes when we calculate the Stokes parameters of the polarization \citep{2014Chang_apj}. We note that the cross-section of the scattering is usually considered to be anisotropic such that $\theta_{\rm o}'$ is included.
The  results of the scattered photons are back-transformed to the PF. First, the cosine of the angle $\mu_{k_{\rm o}'p}' = \hat{\mathbf{k}}_{\rm o}' \cdot \hat{\mathbf{p}}'$ between the scattered photon $\hat{\mathbf{k}}_{\rm o}'$ and the electron $\hat{\mathbf{p}}'$ in the ERF is transformed to the PF angle $\mu_{k_{\rm o}p}$ $$\mu_{k_{\rm o}p}\,=\,\frac{\mu_{k_{\rm o}'p}'+\beta}{1+\mu_{k_{\rm o}'p}'\beta}.$$This result yields the scattered photon's direction $\hat{\mathbf{k}}_{\rm o}$ in the PF $$\hat{\mathbf{k}}_{\rm o}\,=\,\mu_{k_{\rm o}p} \hat{\mathbf{p}}+\sqrt{1-\mu_{k_{\rm o}p}^2} \,\hat{P}[\hat{\mathbf{k}}_{\rm o}',\hat{\mathbf{p}}].$$The corresponding PF frequency $\omega_{\rm o}$ is calculated as $\omega_{\rm o}=\gamma(1+\beta\mu_{k_{\rm o}'p}')\omega_{\rm o}'$. Next, the scattered Stokes vector $\mathbf{s}_{k_{\rm o}'k_{\rm i}'}'$ (from the scattering step) is rotated into a temporary frame $C_{k_{\rm o}'p}' = \{\hat{\mathbf{n}}[\hat{\mathbf{k}}_{\rm o}' \times \hat{\mathbf{p}}], \hat{\mathbf{k}}_{\rm o}'\}$, where the $y$-axis is perpendicular to the $\hat{\mathbf{k}}_{\rm o}'$-$\hat{\mathbf{p}}$ plane: $\mathbf{s}_{k_{\rm o}'p}' = \mathbf{M}[\Delta[C_{k_{\rm o}'k_{\rm i}'},C_{k_{\rm o}'p}]] \mathbf{s}_{k_{\rm o}'k_{\rm i}'}'$. The invariance of the Stokes vector under the Lorentz boost ensures that the transformed vector $\mathbf{s}_{k_{\rm o}p}$ in the PF (relative to $C_{k_{\rm o}p} = \{\hat{\mathbf{n}}[\hat{\mathbf{k}}_{\rm o} \times \hat{\mathbf{p}}], \hat{\mathbf{k}}_{\rm o}\}$) is identical to the ERF result $$\mathbf{s}_{k_{\rm o}p}\,=\,\mathbf{s}_{k_{\rm o}'p}'.$$Finally, the Stokes vector is rotated into the final coordinate system $C_{k_{\rm o}k_{\rm i}} = \{\hat{P}[\hat{\mathbf{k}}_{\rm i},\hat{\mathbf{k}}_{\rm o}],\hat{\mathbf{k}}_{\rm o}\}$, which aligns the $y$-axis with the projection of the incident photon's direction $\hat{\mathbf{k}}_{\rm i}$ onto the plane perpendicular to the final propagation direction $\hat{\mathbf{k}}_{\rm o}$:$$\mathbf{s}_{k_{\rm o}k_{\rm i}}\,=\,\mathbf{M}[\Delta[C_{k_{\rm o}p},C_{k_{\rm o}k_{\rm i}}]]\,\,\mathbf{s}_{k_{\rm o}p}.$$ 

The spectra and polarization are reconstructed by counting scattered photons arriving along a fixed observer direction $\hat{k}_\text{obs}$. For an observer with a viewing cone $\theta_{\text{bin}}$, the contribution in a frequency bin centered at $\omega_{\text{o}}$ with width $\Delta\omega_{\text{o}}$ is obtained by summing the Stokes weights of all photons that satisfy $\mu=\hat{k_\text{o}}\cdot\hat{k}_\text{obs}>\cos\theta_{\text{bin}}$ and $\omega_{\text{o}}\in[\omega_{\text{o}}-\Delta\omega_{\text{o}}/2,\ \omega_{\text{o}}+\Delta\omega_{\text{o}}/2]$. The binned Stokes spectra at a given scattered frequency $\omega_{\rm o}$ are constructed by summing the Stokes weights of all photons in the corresponding angular and frequency bin,
\[
I(\omega_{\rm o}) = \sum_{k=1}^{N}I_k, \qquad
Q(\omega_{\rm o}) = \sum_{k=1}^{N} Q_k, \qquad
U(\omega_{\rm o}) = \sum_{k=1}^{N} U_k,
\]
where the sum runs over all selected photons contributing to that frequency bin. In the present calculations, the incident photons are initialized with a normalized Stokes vector $(I,Q,U)=(1,1,0)$, and each scattered photon carries a unit intensity weight ($I_k=1$). As a result, $I(\omega_{\rm o})$ corresponds to the total number of photons $N$ in the selected direction \citep{matt1996,Dreyer2021ApJ}.
For comparison with band-integrated observational measurements, we further integrate the Stokes parameters over a finite energy band defined by $\omega_1 \le \omega_{\rm o} \le \omega_2$. 
In this case, the total Stokes parameters are obtained by summing over all photons within the band,
\[
I_{\rm tot} = \sum_i I_i, \qquad
Q_{\rm tot} = \sum_i Q_i, \qquad
U_{\rm tot} = \sum_i U_i.
\]
The band-averaged degree of polarization is then computed as $\Pi_{\rm band}=100\times\sqrt{Q_{\rm tot}^2+U_{\rm tot}^2}/I_{\rm tot}(\%)$. To ensure sufficient photon statistics, we simulate $10^{9}$ scattering events and collect photons within a viewing cone of $\pm4^{\circ}$ around the observer direction.

\section{Semi-Analytic Model for the Polarization of IC Emission}\label{appendixb}

Based on the work of \citet{Bonometto1970A&A}, \citet{Krawczynski2012ApJ} derived SA results for the polarization dependence on frequency in IC scattering. Here, we repeat the treatment of \citet{Krawczynski2012ApJ} to illustrate the analytic process of IC polarization. 

The maximum frequency $\omega_{\rm max}$ emitted by an electron with Lorentz factor $\gamma$ moving in the direction $\hat{\mathbf{p}}$ relative to the incident photon $\hat{\mathbf{k}}_{\rm i}$ is $\omega_{\rm max} = 2\gamma^2 (1-\beta \mu_{k_{\rm i}p})\omega_{\rm i}$, where $\mu_{k_{\rm i}p}=\hat{\mathbf{k}}_{\rm i}\cdot\hat{\mathbf{p}}$. From this, the minimum Lorentz factor required to produce a scattered photon of frequency $\omega_{\rm o}$ is approximated (for $\gamma \gg 1$ and $\beta \approx 1$) as $$\gamma_{\rm min} = \sqrt{\frac{1}{2} \frac{\omega_{\rm o}}{\omega_{\rm i}} \frac{1}{(1-\mu_{k_{\rm i}p})}}.$$We introduce the dimensionless frequency variable $y$ as the ratio of the observed frequency $\omega_{\rm o}$ to the maximum frequency emitted by an electron with a factor $\gamma_0$ in that direction:$y = \omega_{\rm o}/\omega_{\rm max}(\gamma_0,\mu_{k_{\rm i}p})$. For monoenergetic electrons ($\gamma_0$), the differential photon number (intensity kernel) $\rho_{\gamma_0}$ and the degree of polarization $\Pi_{\gamma_0}$ are functions of $y$ $$\rho_{\gamma_0} \equiv \frac{dN_{\gamma}}{d\omega_{\rm o}} = \frac{3(1-2 y+2 y^2)}{2 \,\omega_{\rm max}} \quad \text{and} \quad \Pi_{\gamma_0} = \frac{y^2}{1-2 y+2 y^2}.$$These equations serve as the integration kernel for predicting the spectral shape and polarization properties from complex electron distributions.

To predict the spectral shape and polarization properties for an arbitrary electron energy distribution $\frac{dN_{\rm e}}{d\gamma}$, the frequency-dependent degree of polarization $\Pi$ is calculated by integrating the contributions of $\rho_{\gamma_0}$ and $\Pi_{\gamma_0}$ over the electron Lorentz factor $\gamma$ $$\Pi(\omega_{\rm o}) =\Pi_{\rm init} \frac{\int_{\gamma_1'}^{\gamma_2}\, \frac{dN_{\rm e}}{d\gamma}\, \rho_{\gamma_0}\, \Pi_{\gamma_0}\, d\gamma}{\int_{\gamma_1'}^{\gamma_2}\, \frac{dN_{\rm e}}{d\gamma}\, \rho_{\gamma_0} \,d\gamma}$$where the integral is carried out over the energy range $[\gamma_1', \gamma_2]$, and the lower limit $\gamma_1' = \mathrm{Max}(\gamma_1,\gamma_{\rm min})$ ensures that only electrons capable of producing the observed photon frequency contribute, $\Pi_{\rm init}$ denotes the initial degree of polarization of the incident photon. For a given frequency band $(\omega_1,\omega_2)$, the SA
band-averaged polarization is defined as
$$\Pi_{\rm band}=
\frac{\int_{\omega_1}^{\omega_2} I(\omega_{\rm o})\,\Pi(\omega_{\rm o})\,d\omega_{\rm o}}
{\int_{\omega_1}^{\omega_2} I(\omega_{\rm o})\,d\omega_{\rm o}}$$
where $I(\omega_{\rm o})\equiv \omega_{\rm o}^2\,dN/d\omega_{\rm o}$, $\Pi(\omega_{\rm o})$ is the
frequency-dependent polarization obtained from the electron-distribution average.

\section{Validation of the Polarization Calculations}
\begin{figure}
\includegraphics[width=\columnwidth]{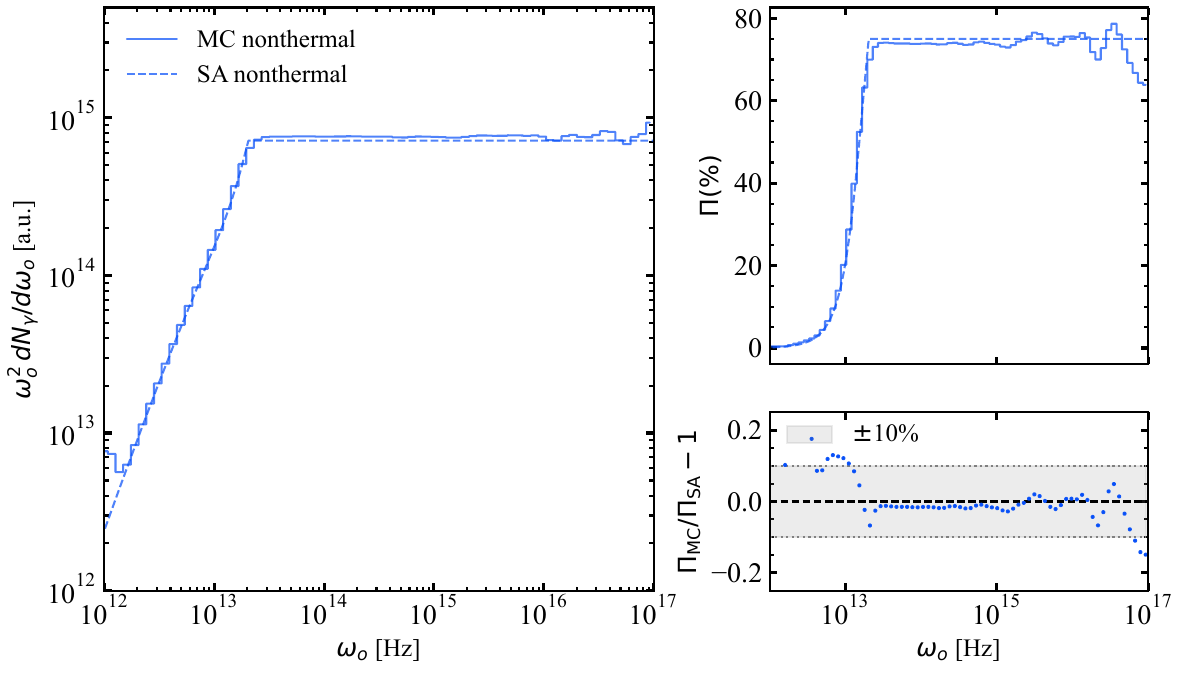}
\caption{Comparison of numerical results for intensity and polarization with the theoretical predictions of \citet{Bonometto1970A&A}. The electron distribution parameters are set to $\gamma_{1} = 60$, $\gamma_{2} = 10^6$, $p_{\text{hybrid}}=3$, and $\gamma_{\mathrm{th}} = 60$. The incident monochromatic beam has a frequency of $\omega = 3.1 \times 10^9$ Hz (corresponding to a dimensionless energy $\epsilon = 2.5 \times 10^{-11}$) and an initial Stokes vector $s = (1, 1, 0)$. The observation direction is set to $(\theta_{\rm o}, \phi_{\rm o})=(85^\circ, 0^\circ)$.}
\label{fig:fig_compare}
\end{figure}

To validate our numerical code, we simulate single Compton scattering of a monochromatic and completely polarized photon beam. The photons are scattered by electrons following a pure power-law distribution ($\gamma_{1} = 60$, $\gamma_{2} = 10^6$, $p_{\text{hybrid}}=3$, $\gamma_\text{th}= 60$; i.e., the entire electron population follows a power-law with no thermal component). The results are shown in Figure \ref{fig:fig_compare}. The parameter set is chosen to be identical to those adopted in previous studies, enabling a direct and meaningful comparison with the results of \citet{Moscibrodzka2022ApJS} (see their Figure 2) and \citet{Yang2021ApJS} (see their Figure 25). Our results are consistent with those obtained in both studies, which were obtained using independent numerical schemes.

\section{Additional materials}
\begin{figure*}
\centering
\begin{tabular}{@{}cc@{}}
\includegraphics[width=0.49\textwidth]{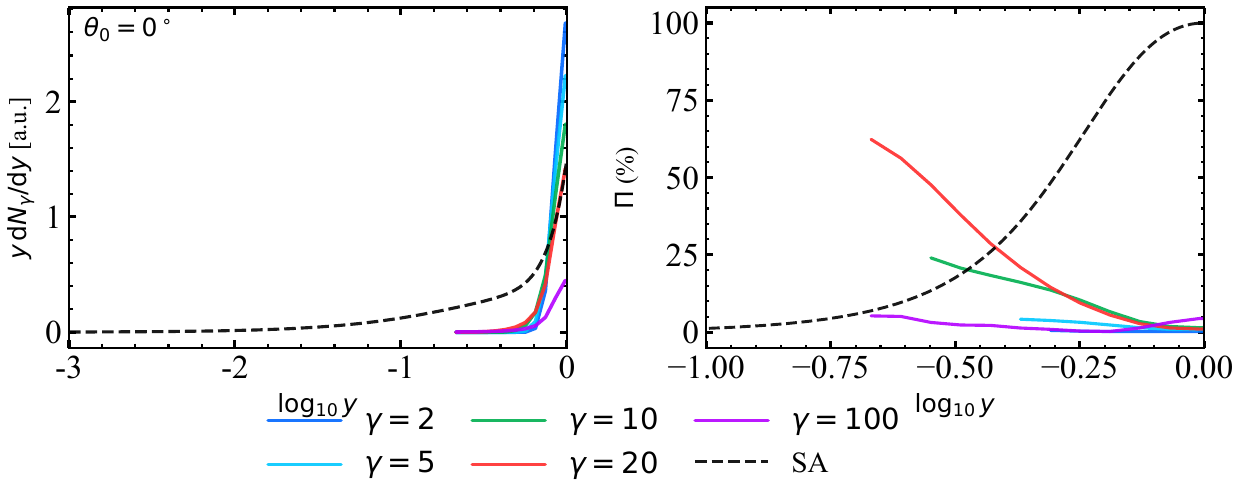} &
\includegraphics[width=0.49\textwidth]{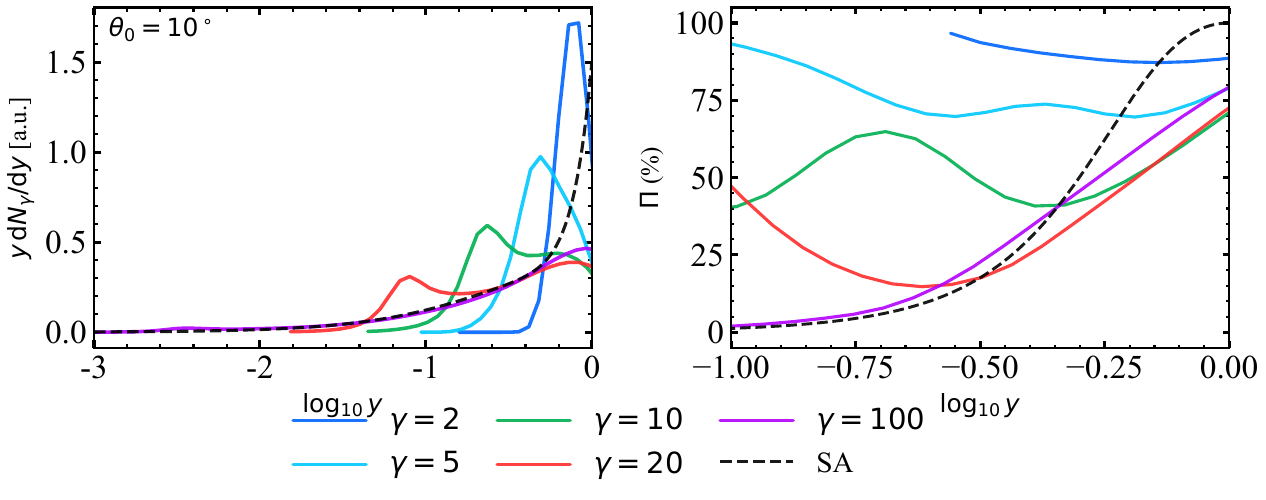} \\

\includegraphics[width=0.49\textwidth]{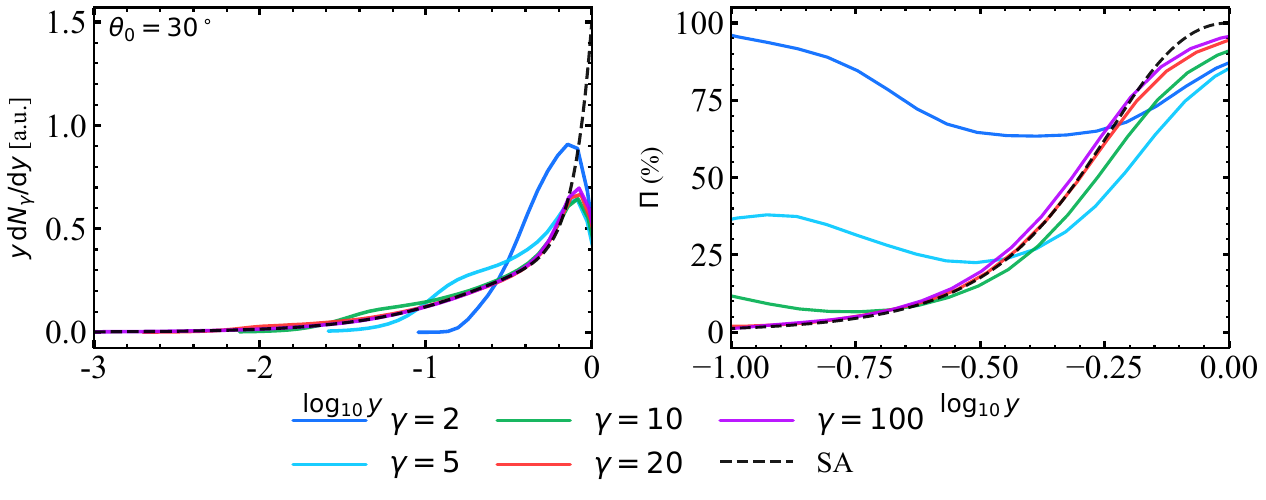} &
\includegraphics[width=0.49\textwidth]{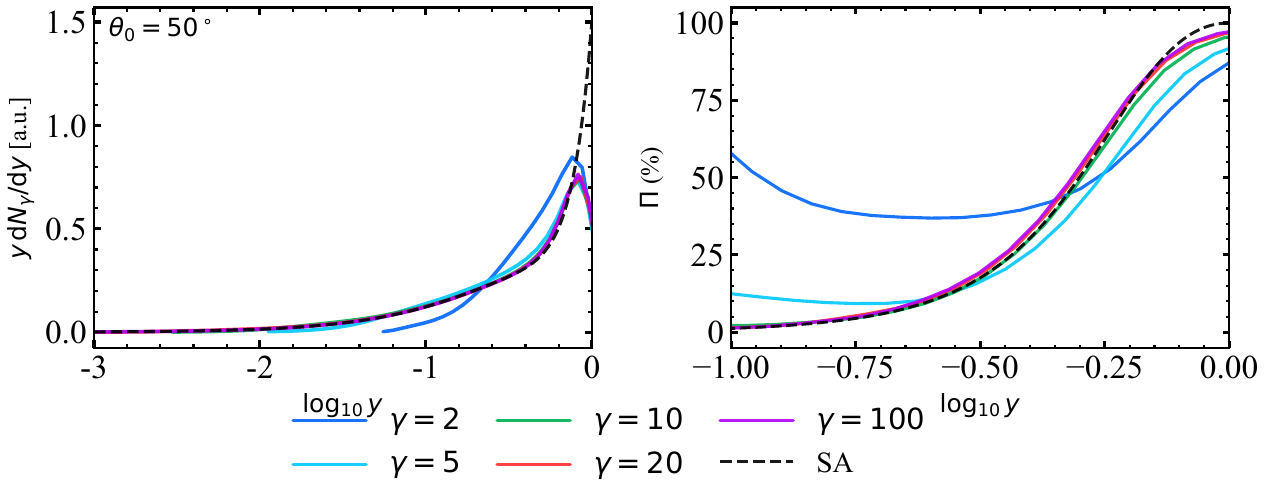} \\

\includegraphics[width=0.49\textwidth]{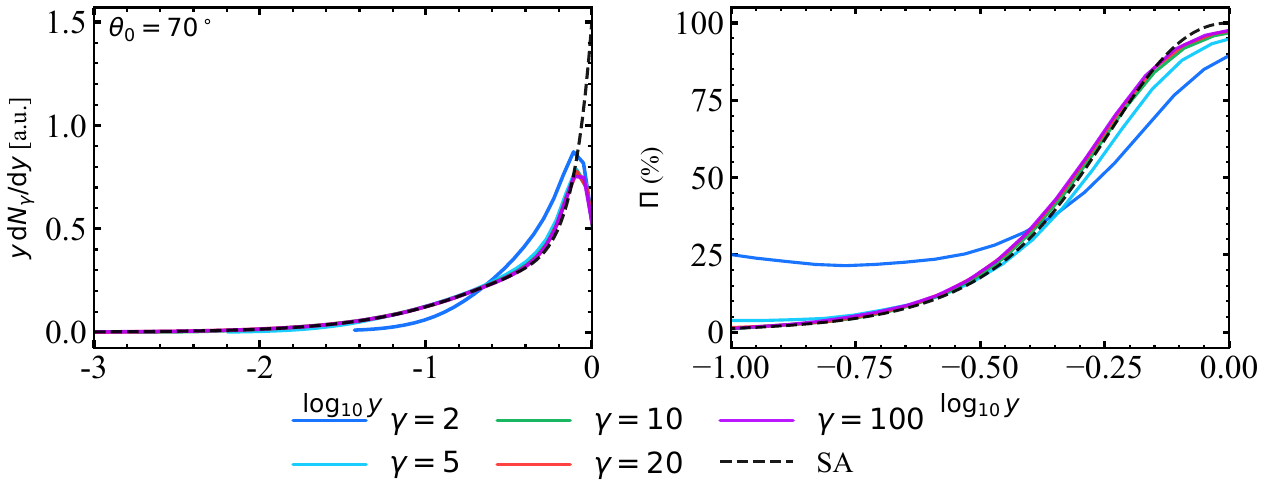}&
\includegraphics[width=0.49\textwidth]{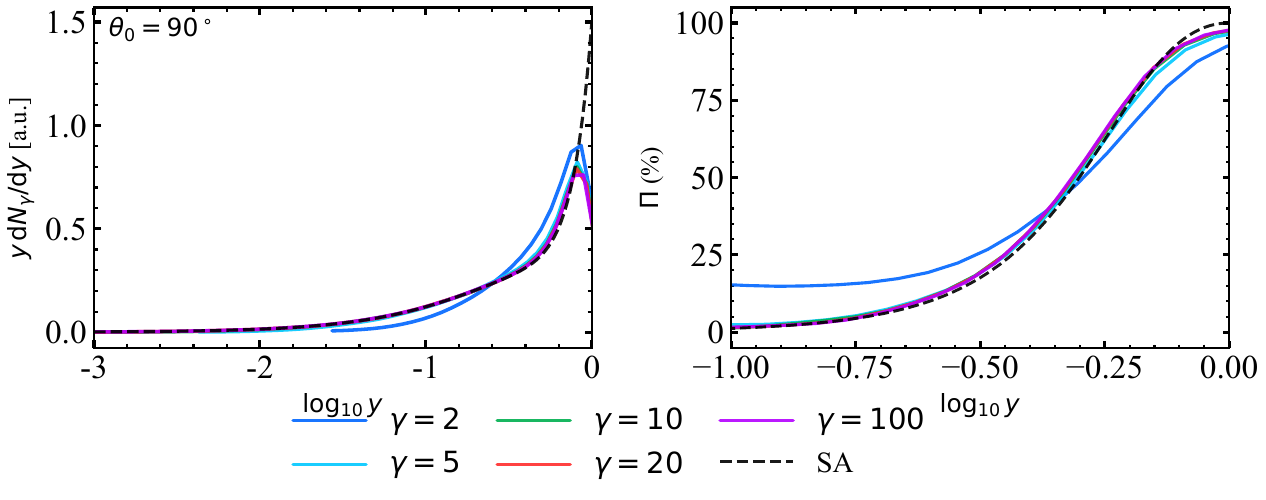}
\end{tabular}

\caption{Same as Figure~\ref{fig:fig4}, but for the viewing angles $\theta_{\rm o}$ are $0^\circ$, $10^\circ$, $30^\circ$, $50^\circ$, $70^\circ$ and $90^\circ$.}
\label{fig:fig4_app_1}
\end{figure*}

\begin{figure*}
\centering
\begin{tabular}{@{}cc@{}}
\includegraphics[width=0.49\textwidth]{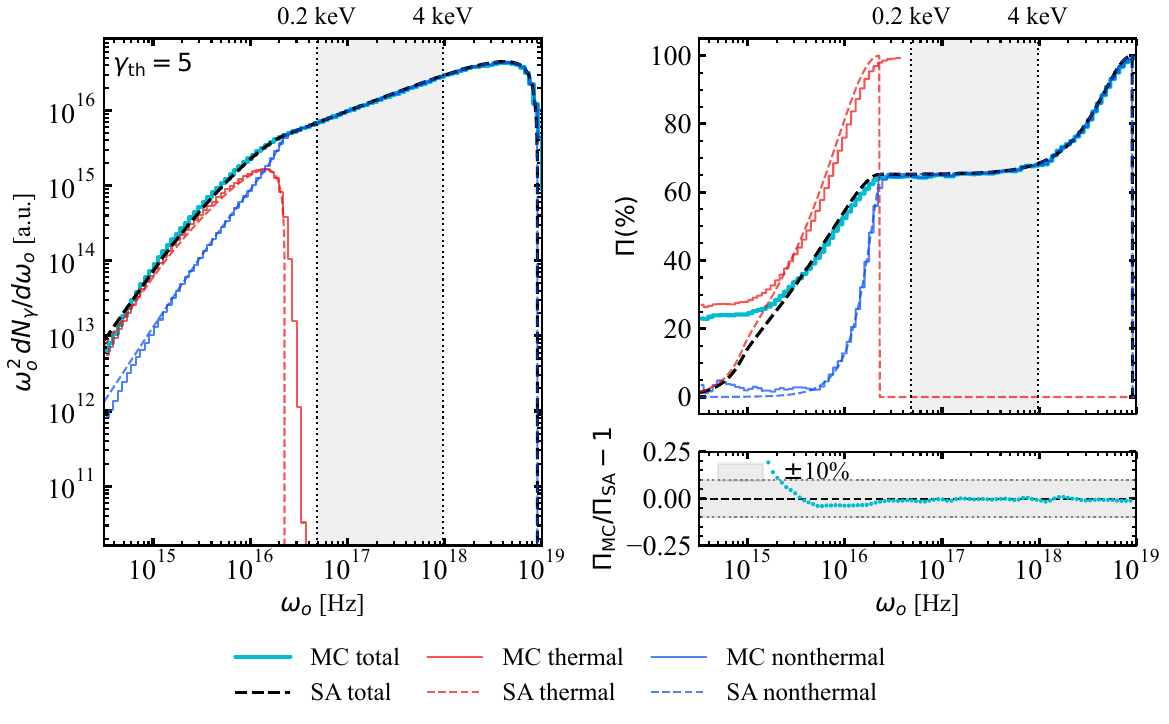} &
\includegraphics[width=0.49\textwidth]{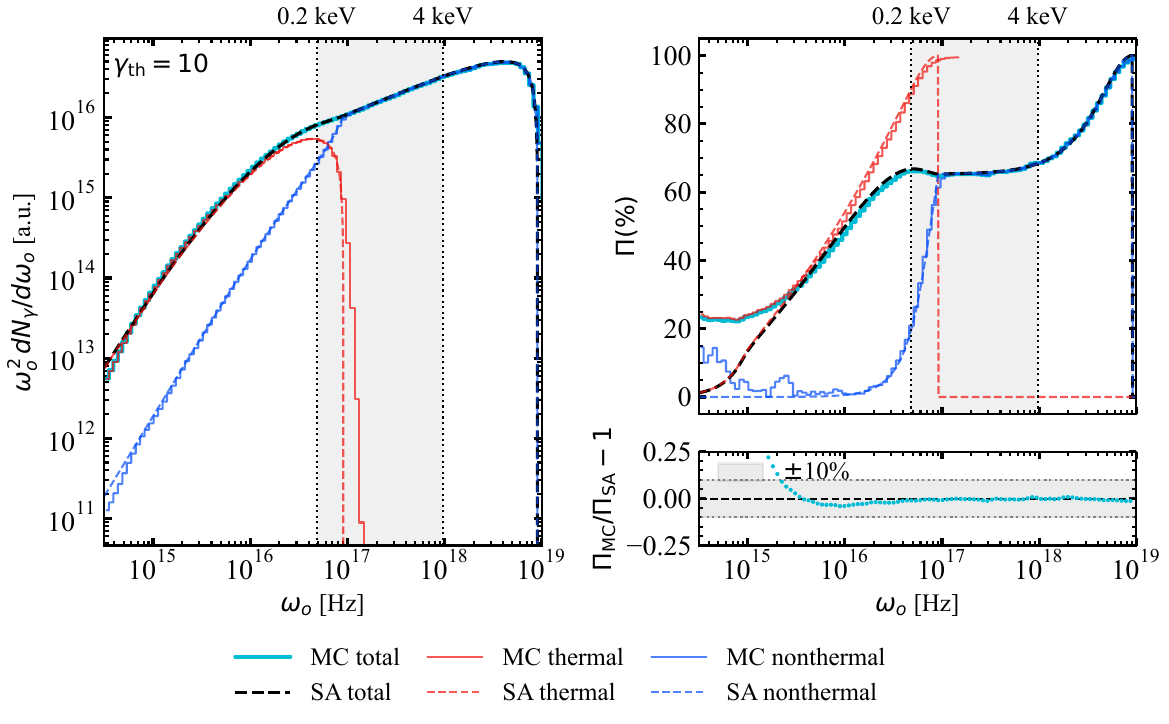} \\

\includegraphics[width=0.49\textwidth]{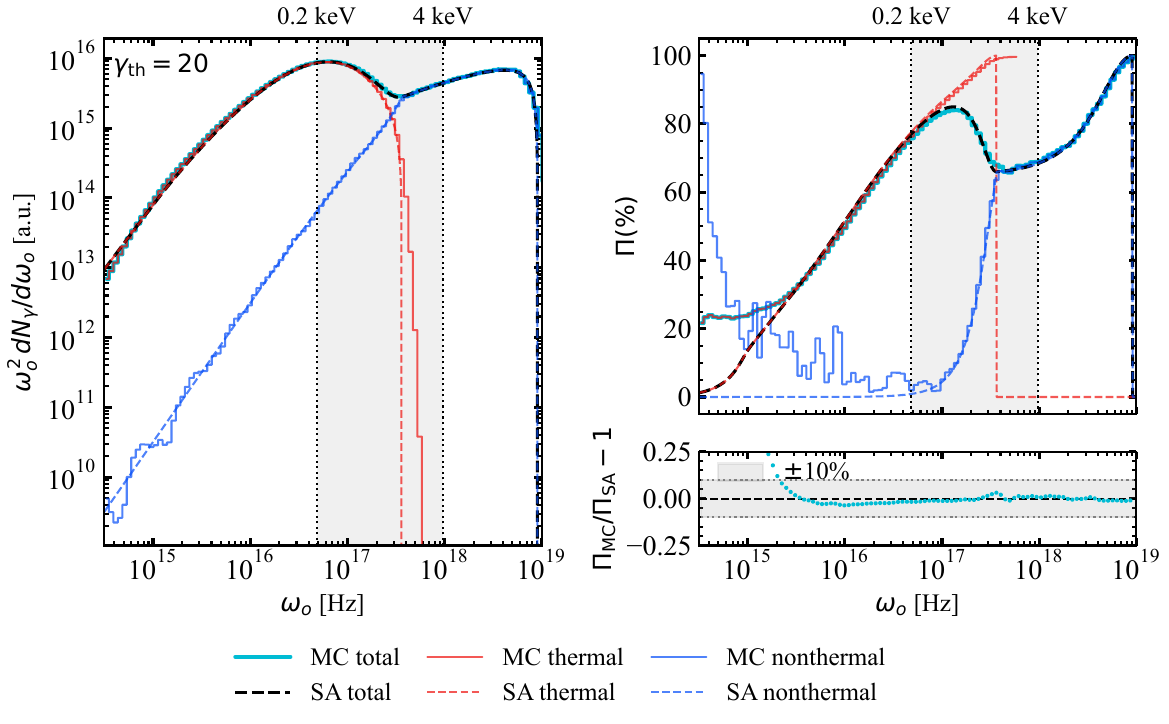} &
\includegraphics[width=0.49\textwidth]{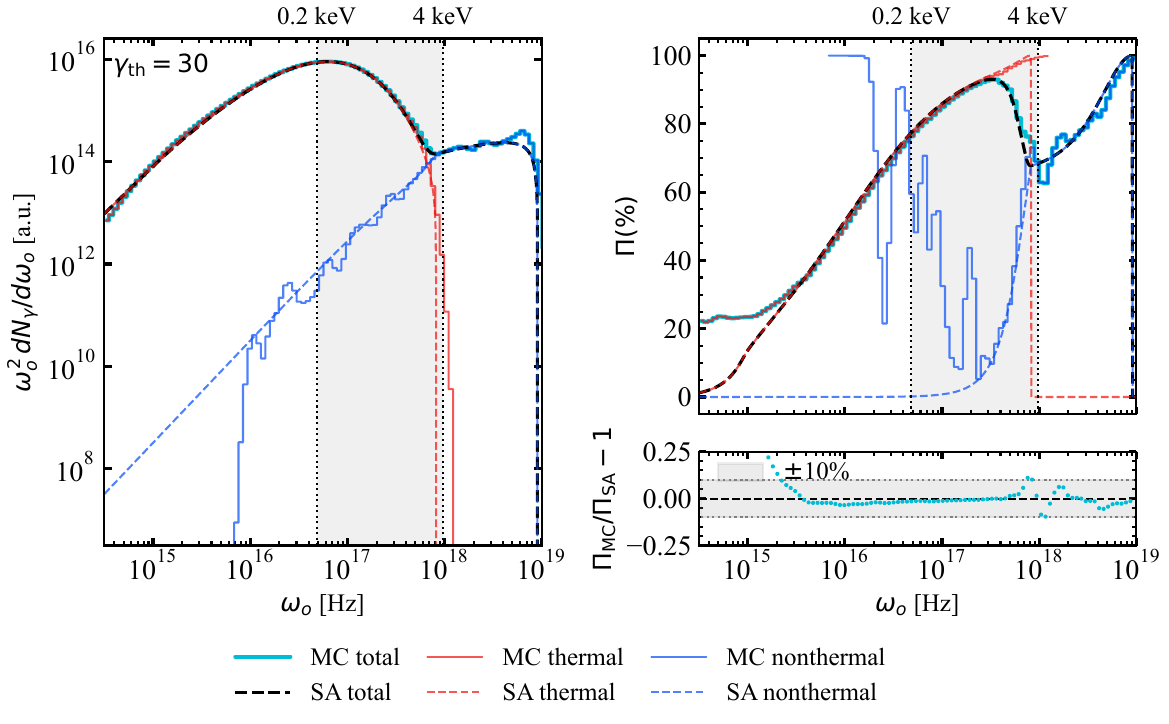} 

\end{tabular}

\caption{Same as Figure~\ref{fig:fig7}(a), but for different values of the transition Lorentz factor $\gamma_{\rm th}$ for 5, 10, 20, and 30.}
\label{fig:fig7_app_1}
\end{figure*}

\begin{figure*}
\centering
\begin{tabular}{@{}cc@{}}
\includegraphics[width=0.42\textwidth]{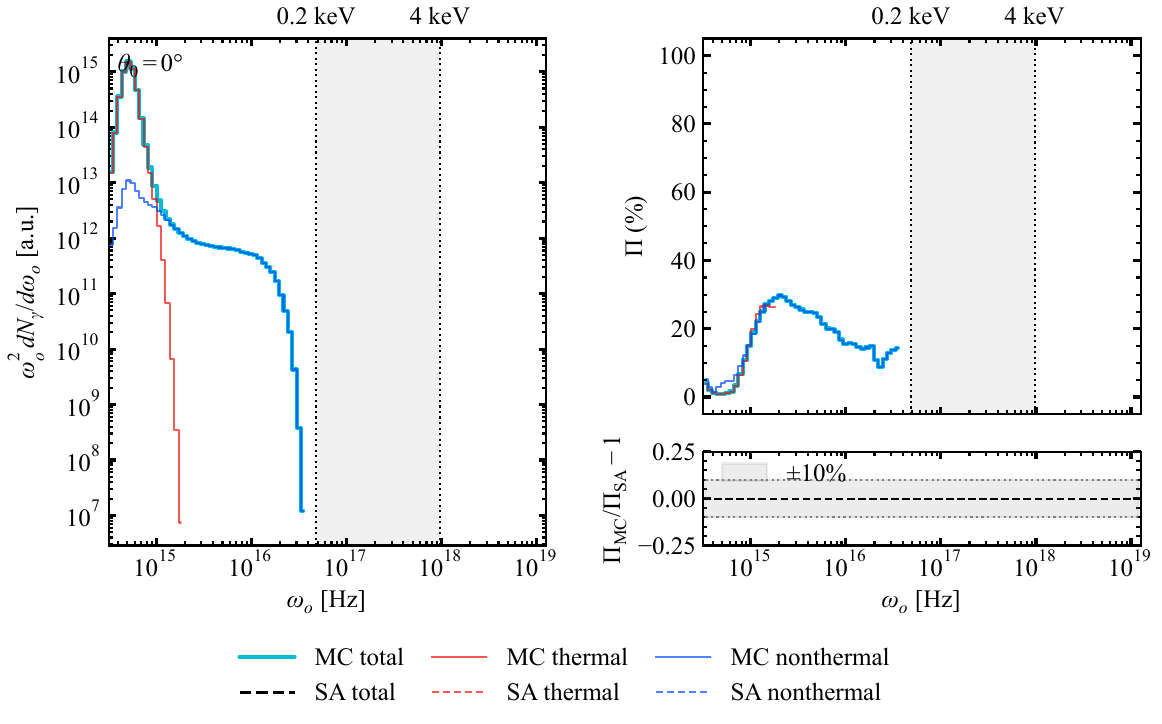} &
\includegraphics[width=0.42\textwidth]{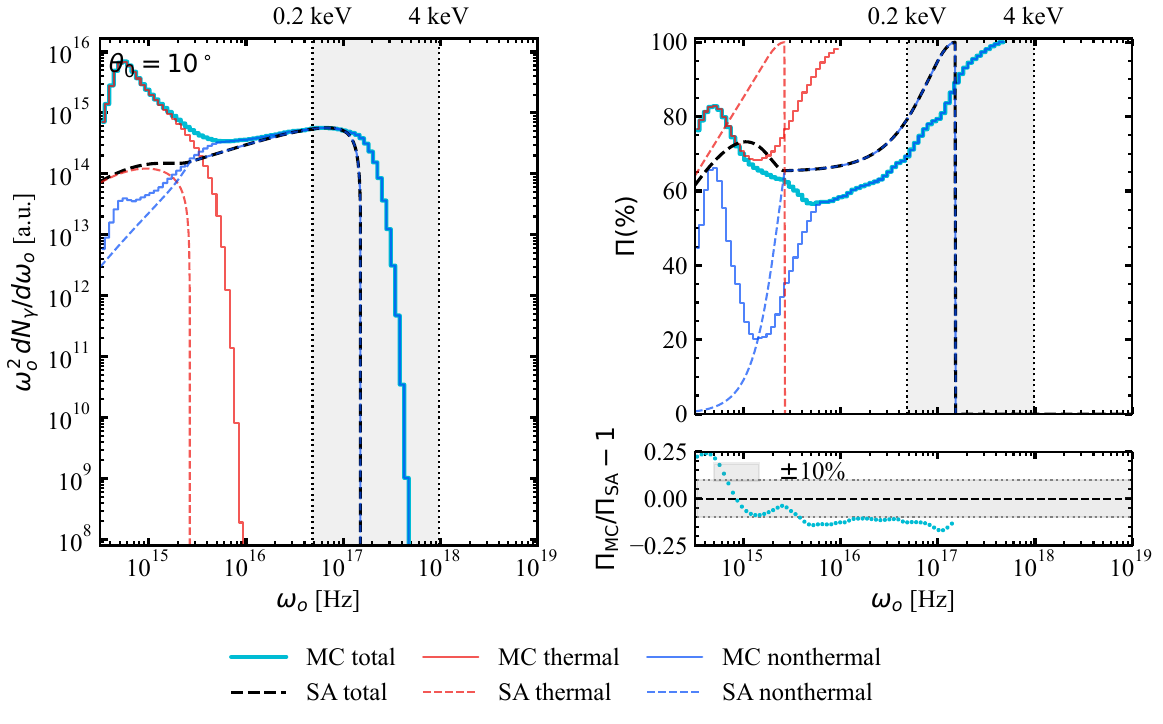} \\

\includegraphics[width=0.42\textwidth]{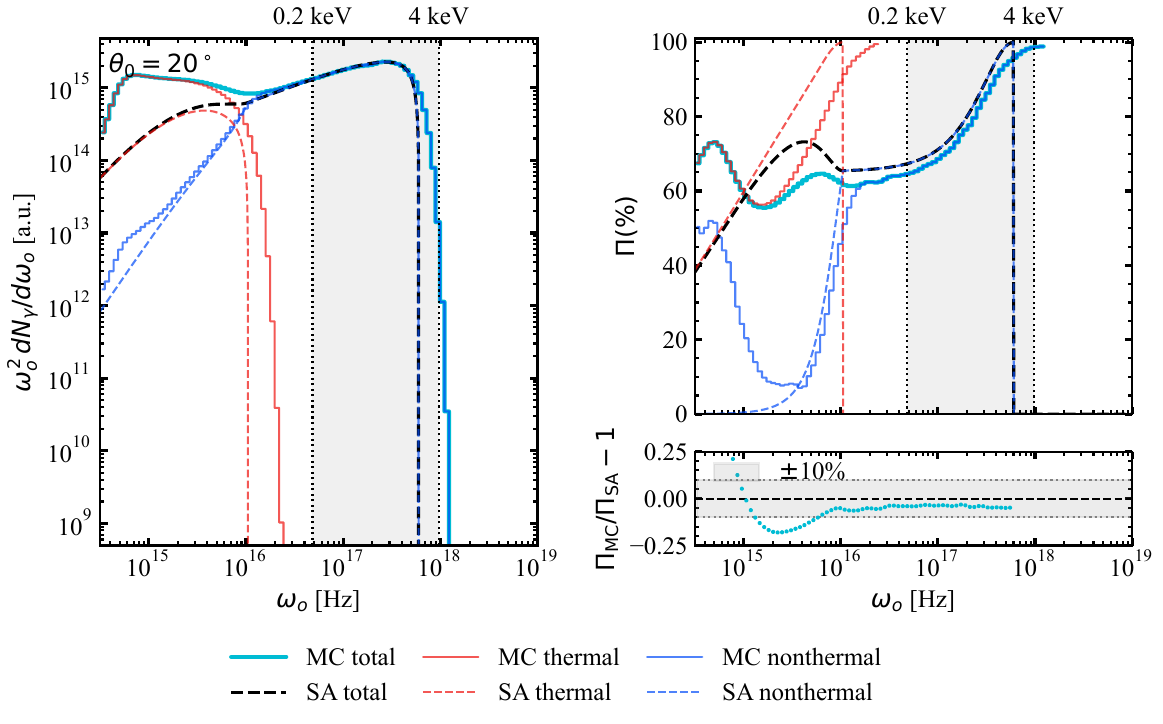} &
\includegraphics[width=0.42\textwidth]{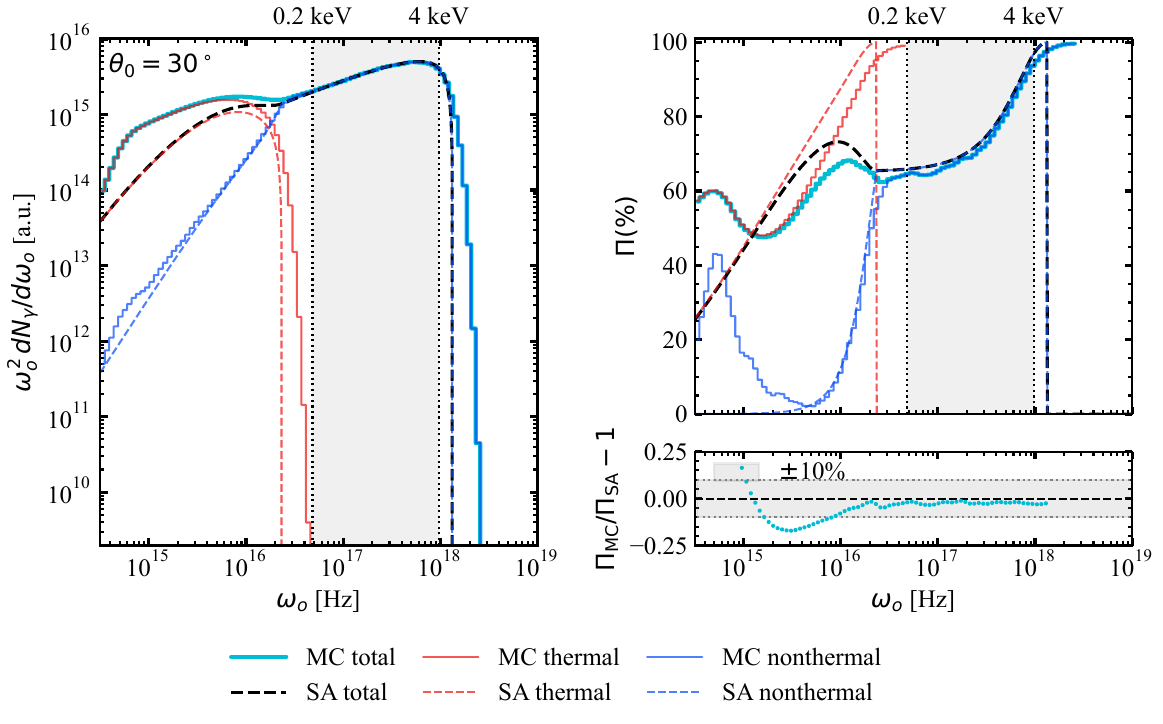} \\

\includegraphics[width=0.42\textwidth]{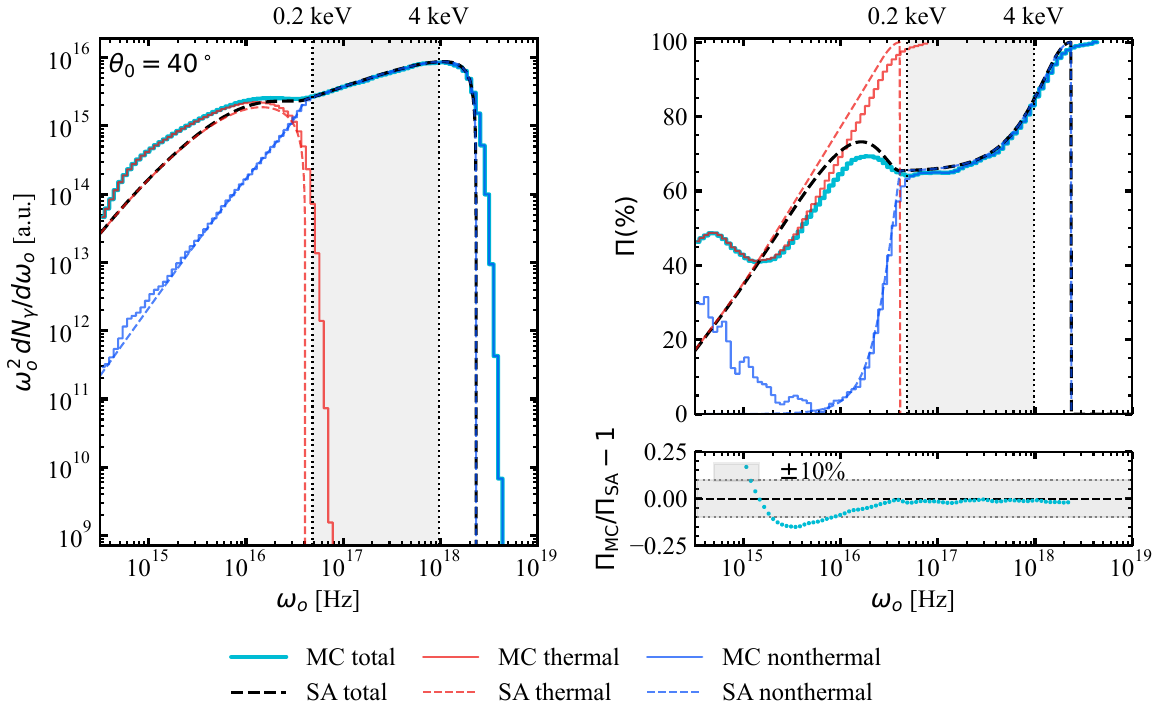} &
\includegraphics[width=0.42\textwidth]{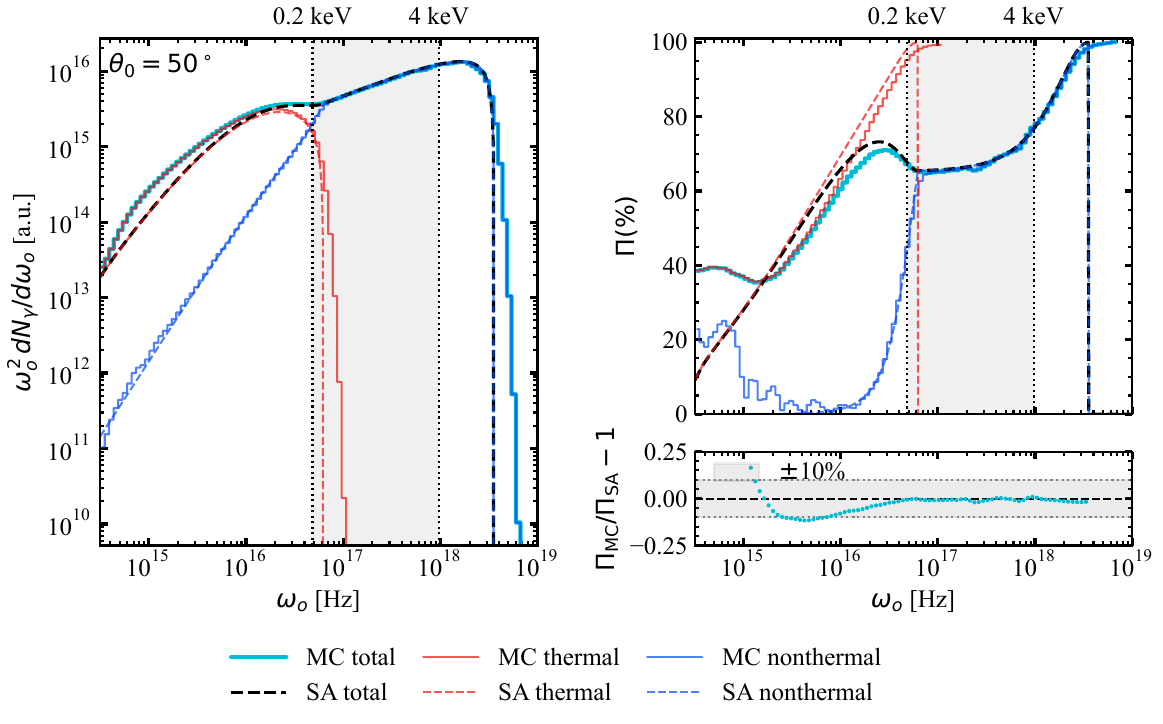} \\

\includegraphics[width=0.42\textwidth]{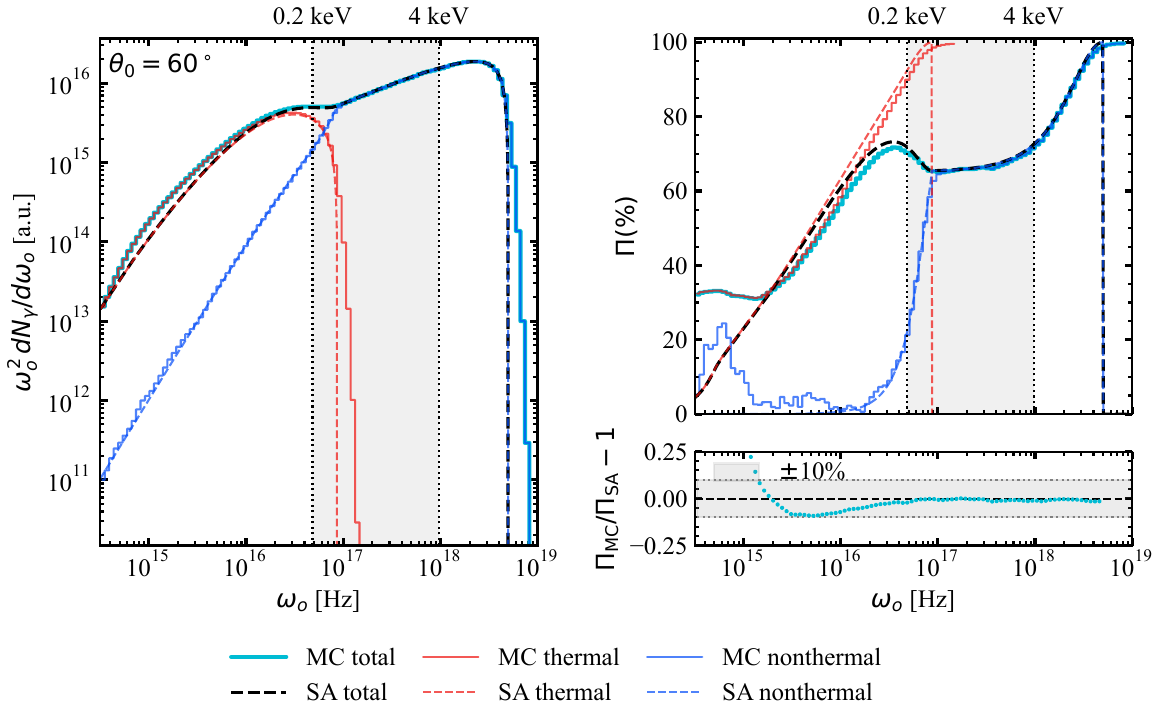} &
\includegraphics[width=0.42\textwidth]{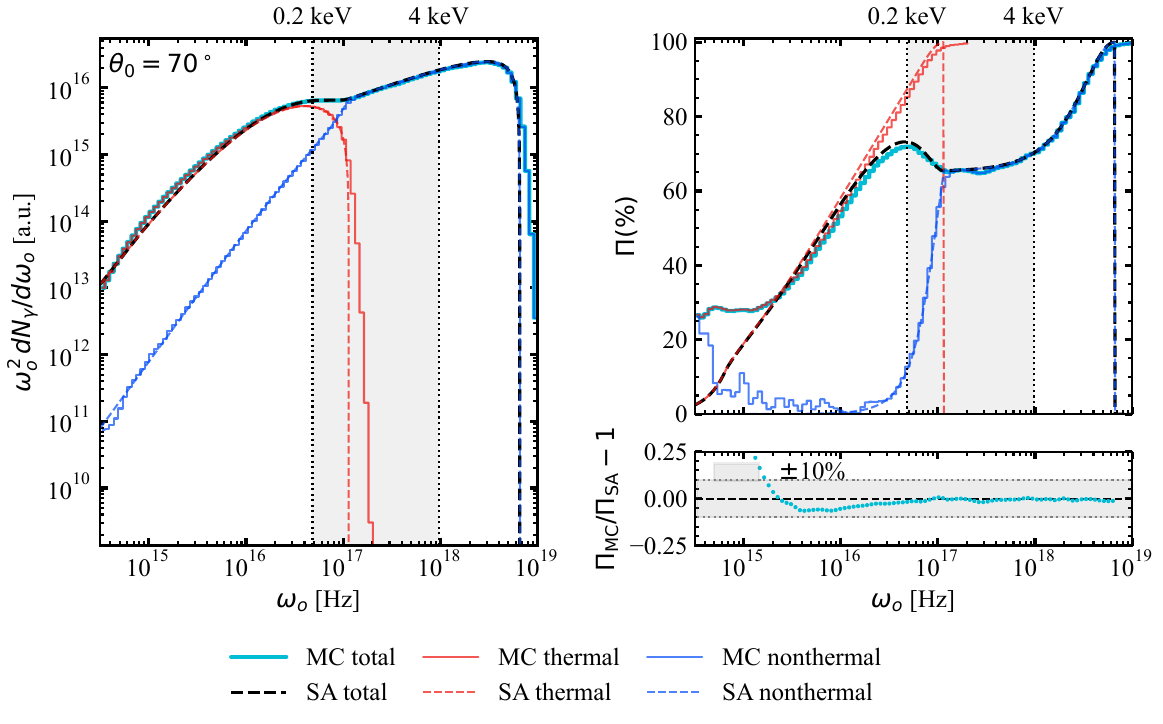} \\

\includegraphics[width=0.42\textwidth]{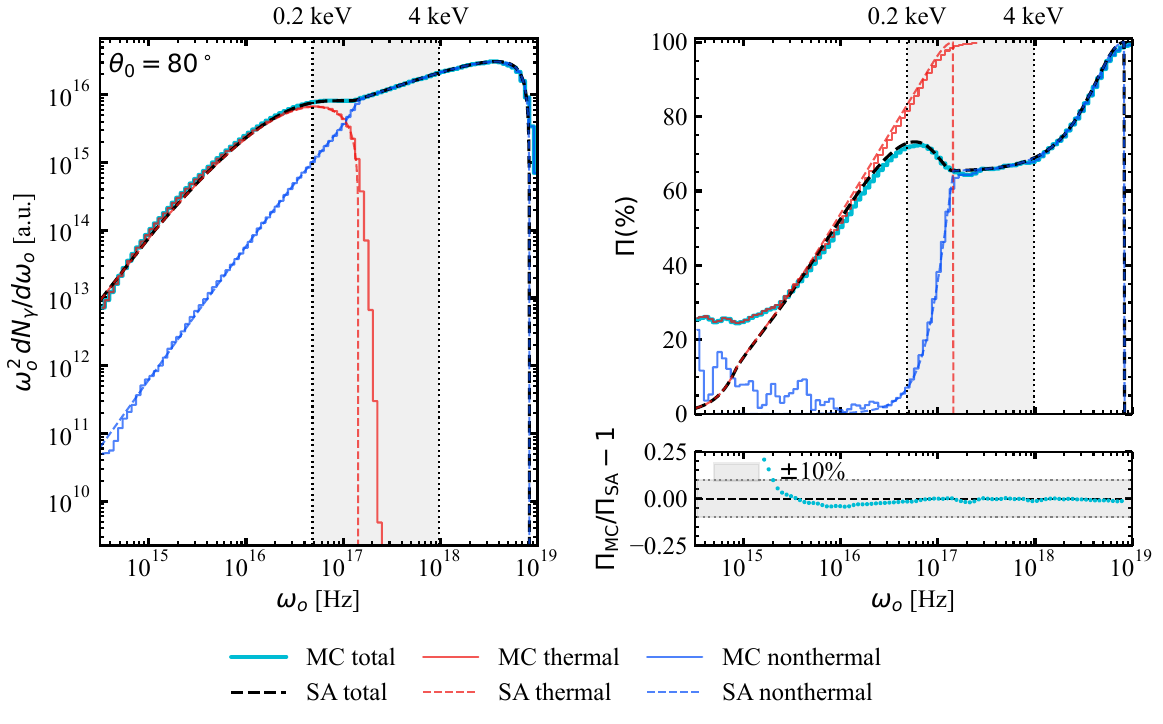} &
\includegraphics[width=0.42\textwidth]{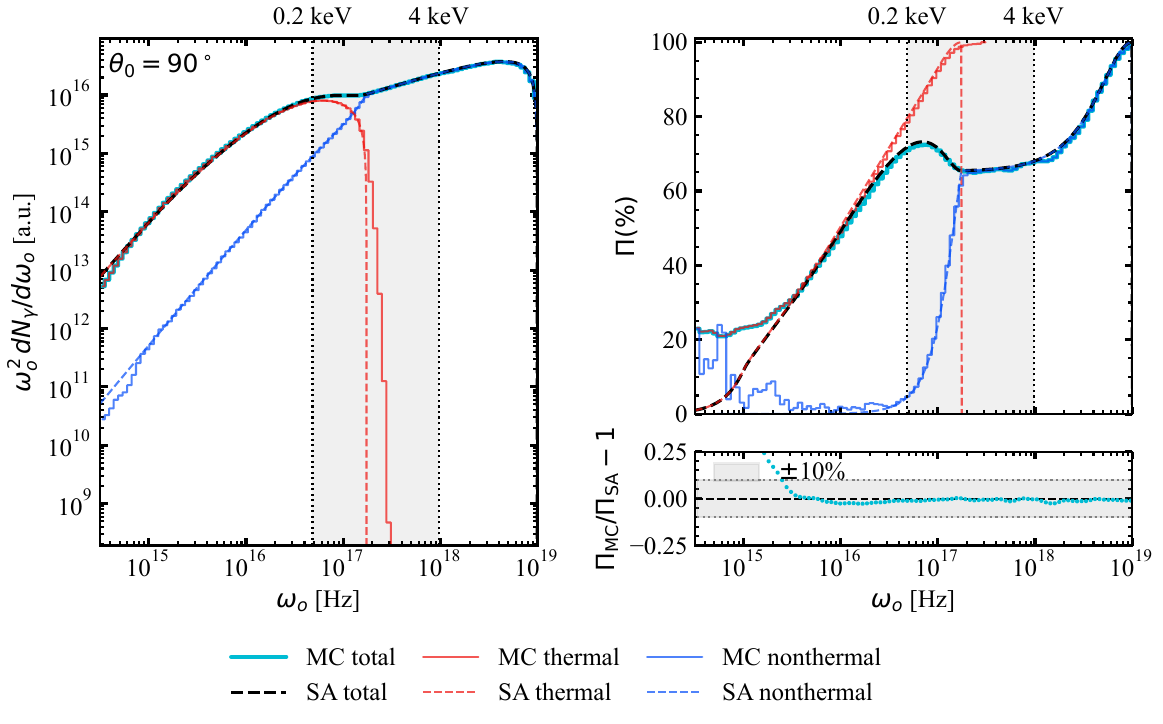}
\end{tabular}

\caption{Same as Figure~\ref{fig:fig7}(a), but for different values of the viewing angle $\theta_{\rm o}$ in the range from 0 to 90.}
\label{fig:fig8_app_1}
\end{figure*}

Here we list the results for different viewing angles for the scattering of monoenergetic photons by isotropic, 
monoenergetic electrons, as presented in Figure \ref{fig:fig4_app_1}.
We also list the corresponding frequency-dependent intensity and polarization spectra in the hybrid electron energy 
distribution for different values of $\gamma_{\rm th}$, as presented in Figure \ref{fig:fig7_app_1}.
Finally, we list the corresponding frequency-dependent intensity and polarization spectra in the hybrid electron 
energy distribution for different viewing angles $\theta_{\rm o}$, as presented in Figure \ref{fig:fig8_app_1}.

\label{lastpage}

\end{document}